\documentclass[10pt]{article}
\usepackage{cite}
\usepackage{ulem}
\begin{document}
\begin{titlepage}
\title{Black Strings, Black Rings and State-space Manifold}
\author{}
\date{Stefano Bellucci $^{a}$ \thanks{\noindent bellucci@lnf.infn.it} \ and
 Bhupendra Nath Tiwari $^{a}$ \thanks{\noindent bntiwari.iitk@gmail.com}\\
\vspace{0.5cm}
$^{a}$ INFN-Laboratori Nazionali di Frascati\\
Via E. Fermi 40, 00044 Frascati, Italy.} \maketitle
\abstract{State-space geometry is considered, for diverse three
and four parameter non-spherical horizon rotating black brane
configurations, in string theory and $M$-theory. We have
explicitly examined the case of unit Kaluza-Klein momentum
$D_1D_5P$ black strings, circular strings, small black rings and
black supertubes. An investigation of the state-space pair
correlation functions shows that there exist two classes of brane
statistical configurations, {\it viz.}, the first category
divulges a degenerate intrinsic equilibrium basis, while the
second yields a non-degenerate, curved, intrinsic Riemannian
geometry. Specifically, the solutions with finitely many branes
expose that the two charged rotating $D_1D_5$ black strings and
three charged rotating small black rings consort real degenerate
state-space manifolds. Interestingly, arbitrary valued
$M_5$-dipole charged rotating circular strings and Maldacena
Strominger Witten black rings exhibit non-degenerate, positively
curved, comprehensively regular state-space configurations.
Furthermore, the state-space geometry of single bubbled rings
admits a well-defined, positive definite, everywhere regular and
curved intrinsic Riemannian manifold; except for the two finite
values of conserved electric charge. We also discuss the
implication and potential significance of this work for the
physics of black holes in string theory.}
%$D$-brane charge= electric charge
\vspace{1.5cm}

\textbf{Keywords: Rotating Black Branes; Microscopic Configurations; State-space Geometry.} \\

PACS numbers: 04.70.-s Physics of black holes; 04.70.Bw Classical black holes;
04.70.Dy Quantum aspects of black holes, evaporation, thermodynamics.
\end{titlepage}

%\newpage
\begin{Large} \textbf{Contents:} \end{Large}\\
\begin{enumerate}
\item{Introduction.}
\item{$\mathcal N=1$ Supergravity Origin.}
\item{State-space Geometry.}
\item{Black Rings and Strings:}
\subitem{4.1 \ \ $D_1D_5P$ Black Strings} \subitem{4.2 \ \ Black
Rings as Circular Strings} \subitem{4.3 \ \ Small Black Rings.}
\subitem{4.4 \ \ Supertubes.}
\item{Conclusion and Discussion.}
\item{Future Directions and Open Issues.}
\end{enumerate}
\section{Introduction}
Recently, we have analyzed the state-space configurations of
exactly fluctuating $1/2$-BPS configuration \cite{BNTBull08} and
subsequently extended the state-space analysis for a class of
supersymmetric spherical horizon black holes \cite{BNTBullcorr,
RotBHs} which preserve two real supersymmetries \cite{FHM}. A set
of interesting supergravity configurations \cite{SF1, SF2, SF3}
provide further definite realizations for the present equilibrium
statistical investigations. Several akin solutions also have been
discovered in 5-dimensional gauged supergravity configurations
\cite{gu-re-1,gu-re-2,cclp1,cclp2,klr}. In particular, it has been
realized since the study of BMPV black holes that the $D=5$ vacuum
Einstein equations admit an asymptotically flat, stationary black
hole solution with an event horizon of $S^1 \times S^2$ topology
which describes a rotating black ring \cite{0110260v2}. It turns
out therefore that the state-space geometry of such a solution
involves finitely many dissolved flux charges. The underlying
argument lies on the fact that the ring solutions may not uniquely
be determined by their conserved charges, mass and angular
momentum, but it rather involves certain dipole charges. Moreover,
we find that the asymptotic charges do not distinguish black rings
from the black holes of spherical topology, and their intrinsic
nature is intimately encoded in the state-space invariant objects.
See for example, the solutions constructed in \cite{HE,EE}
describing an intriguing set of charged black rings. It thus
appears natural to investigate the nature of state-space
quantities of the horizon changing black hole solutions. We shall
determine possible functional relations for the state-space
configuration of black strings and black rings and compare
concerned physical meanings of the global quantities with their
spherical horizon counterparts.

Such a construction would clearly elucidate an important issue of
the state-space geometry of the black rings of \cite{HE,EE}, which
entails the finite violation of the black hole uniqueness
\cite{0110260v2}, since there are finitely many space-time
solutions with the same conserved physical charges. Physically,
one finds that these configurations describe rotating loops of
magnetically charged black strings. Since the loop is
contractible, hence these solutions carry no net magnetic charge.
However, these ring solutions do carry a non-zero dipole moment,
which is a non-conserved quantity, often referred to as magnetic
dipole charge, and thus we explore its plausible entanglement with
underlying state-space geometry. In this case, the black rings of
\cite{RE} correspond to the case of rotating circular strings,
whose state-space geometry may easily be characterized by the
mass, angular momenta, and a dipole charge. We find in general
that the state-space geometric quantities thus obtained, in terms
of a set of given conserved charge(s), dipole charge(s) and
angular momenta, account for the underlying structure of the
continuous infinity of five dimensional black brane
configurations.

Furthermore, the dipole black ring charges have a simple
microscopic interpretation \cite{RE}, which may be obtained from
the dimensional reduction of the eleven dimensional supergravity
solution describing $M_5$-branes, with four worldvolume directions
wrapped on an internal six torus and one worldvolume direction
forming the $S^1$ of the black ring in the non compact dimensions.
One thus finds that the dipole charge of black ring is nothing
more than the number of $M_5$-branes present in the ring
configuration. The most general solution of \cite{RE} has the
three independent dipole charges, and thus describes a $(n+3)$
dimensional state-space manifold. The existence of such a
configuration follows from the well known fact that the dipole
charges arise from the orthogonal intersection of three stacks of
$M_5$-branes wrapped on $T^6$, with the common strings of
intersection forming the $S^1$ of the ring. Classically, the
dipole charges contributing non-trivially to the state-space
quantities are continuous parameters, whereas in the quantum
theory they are quantified in terms of the number of branes in
each stack.

The conserved charges of supersymmetric black ring configurations
may in fact be made arbitrarily close to those of the BMPV black
hole by taking the small radius limit, but the ring cannot carry
exactly the same conserved charges as that of the BMPV black hole.
This is because dipole charges are required to specify the ring
configuration, and subsequently entail the continuous violation of
the four dimensional black hole uniqueness theorem
\cite{0110260v2,HE,EE}. This article shows in characteristic that
the state-space geometry of a ring configuration is relatively
higher dimensional than usual BMPV black holes. Similar
conclusions follow for the non-supersymmetric dipole rings which
are prone to have definite phase transitions as well, and exactly
in the same way as the supersymmetric black rings, they possess a
relatively higher dimensional state-space manifold. However, if
one takes charge quantizations into account, then this violation
of uniqueness in both cases is rendered to be finite, see for
details \cite{EEMR2}. The present consideration confronts with the
case of very large charge configurations and treats the underlying
brane charges as real variables.

Moreover, underlying angular momenta, along with the dipole
momenta, play an important role in understanding the state-space
geometry of yet an another class of higher dimensional black brane
solutions. The recent solutions of supertubes \cite{MT,EMT,MNT}
thus be our upright interest, whose analysis finds firm ground
from the motivation of known state-space constructions
\cite{RotBHs}. Here, the black ring configurations become black
tubes, when lifted to higher dimensions. Ref. \cite{EE} identifies
certain charged non-supersymmetric black tubes as thermally
excited states of two-charge supertubes carrying $D_1$ and
$D_5$-brane charges and a dipole charge associated to the
Kaluza-Klein monopole (KKM), which correspondingly parameterize a
three dimensional state-space geometry in the limiting picture of
given rotation. Supertubes have further been the subject of an
intense interest from yet a different direction, following the
realization that the non-singular, horizon-free supergravity
solutions describing tabular objects are thus in one to one
correspondence with the Ramond sector ground states of the
supersymmetric $D_1$-$D_5$ string intersections
\cite{mathur1,mathurdual,LMM}. In this regard, Mathur has
conjectured that the supergravity solutions describing
three-charge supertubes might similarly be accounted against the
microstates of five dimensional supersymmetric black holes
\cite{mathur2}. Truthfully, Mathur's proposal has motivated a
number of interesting studies for the $D_1$-$D_5$ systems and
supertubes
\cite{GiustoIP,GiustoID,lunin,benakraus,bena,pa-mar,bho,bhko}.
Here, our construction of the state-space geometry examines their
stability criterion, and concerns with an investigation of the
statistical interactions which include the first supergravity
example of non-singular horizon free three-charge supertubes of
\cite{GiustoIP,GiustoID,lunin}.

The state-space investigations and the relationship among black
rings, circular strings and supertubes \cite{0110260v2,HE,EE} may
further be accomplished in the framework of supergravity
solutions. Notice for instance that these configurations do not
admit a supersymmetric limit with an event horizon, and this
complicates the basic understanding of the microscopic origin of
their entropy \cite{RE}. However, there has been some progress
made in this direction by studying a non-supersymmetric extremal
ring with a non-vanishing horizon possessing leading order
contribution. It has been conjectured, though, that the
supersymmetric black rings should exist in this limit as well
\cite{benakraus,bena}. Ipso facto, one may easily find that the
conclusions drawn from our state-space computations are in perfect
accordance with various additional ingredients of supersymmetry.
The promising conclusions are important due to the following two
reasons. Firstly, many of the well-known black ring solutions of
\cite{0110260v2,HE,EE,RE} are classically believed to be unstable,
whereas a supersymmetric black ring configuration should be
stable. Secondly, it should be possible to facilitate a precise
quantitative comparison between black rings, worldvolume
supertubes, and the microscopic conformal field theory of the
$D_1$-$D_5$ system.

We have thus been motivated to analyze these view-points from the
perspective of their state-space geometry, and in particular focus
our attention on the two-charge supertubes which have originally
been discovered as the solutions of Dirac-Born-Infeld effective
action \cite{0206126} of an illustrious $D$-brane configuration in
Minkowski vacuum \cite{MT}. In the worldvolume picture, the branes
associated to net charges are represented by the fluxes on the
worldvolume of a tubular, higher-dimensional branes; while the
latter description carries no net charge itself but only a dipole
charge. In this description, the back-reactions on the space-time
of branes have however been so far neglected \cite{EMT,MNT}. It
appears from this analysis that the associated state-space
geometry describes the statistical pair correlations, correlation
volume, and in forthright, it indicates the nature of the phase
transitions, if any, underlying these higher-dimensional
supergravity configurations. The latter are thus prone to be
affected by radiative corrections, {\it viz.}, $\alpha^{\prime}
$-corrections, multipole corrections, etc. Moreover, the present
analysis signifies that the effects thus required, in order to
procure the inconsistency between the two descriptions of the
supertube configurations, might be viewed as refinements to the
additional state-space co-ordinate, whose general transformations
entangle with their intrinsic manifolds.

It is worth to mention that the state-space description has
actually provided an extremely illuminating understanding for the
physics of two-charge worldvolume supertubes. In turn, Ref.
\cite{pa-mar,bho,bhko} have led to a new way of analyzing the
counting entropy of the $D_1$-$D_5$ system, which does not utilize
its CFT description. It is therefore desirable to have an
analogous description for the three-charge supertube
configurations, as well. However, the worldvolume description
based on D-branes that incorporates the third dipole charge seems
problematic, since it necessarily corresponds to an object which
cannot be captured by the corresponding open string description,
such as $NS_5$-branes or the Kaluza-Klein monopoles, see for
instance \cite{benakraus} and references therein. Essentially,
this difficulty can be circumvented by going to $M$-theory, where
the three branes with net charges may be taken as three orthogonal
$M_2$-branes, whereas the three dipoles are associated to three
$M_5$-branes. This in turn is the most symmetric realization of
the three charge supertubes, whose statistical configurations may
thus easily be analyzed from the perspective of our state-space
geometry.

Furthermore, Ref. \cite{EEMR2} shows, in the lieu of the present
investigation, that there exist concrete supersymmetric solutions,
arising from the effective action of a single $M_5$-brane in the
$M$-theory Minkowski vacuum \cite{MT}, and that these
configurations may at most carry up to four $M_2$-brane charges
and six $M_5$-brane dipoles. In a generic set up, the state-space
geometry turns out to be rather more involved and the effective
supergravity solutions render to be calibrated supertubes, because
the worldspace of $M_5$-branes takes the form $S \times C$, where
$S$ is a calibrated surface and $C$ is an arbitrary curve living
on the moduli space. Moreover, the three-charge calibrated
supertube configurations capture all the dipoles, and one finds in
this case that an arbitrary cross-section is possible for
consistent supergravity solutions. It thus turns out, as expected
from standard arguments orginating from \cite{benakraus}, that
these solutions in general suffer from certain limitations, and
accordingly we notice that they capture only one of the angular
momenta, as opposed to the two angular momenta present in the
respective supergravity description. Here our goal is thus to
provide a detailed discussion of the state-space geometry
originating from the fundamental string theory and $M$-theory
configurations and thereby to explore the related issues, which we
shall present in the corresponding section of bubbled supertubes.

The present paper concentrates on the macroscopic-microscopic
aspects of non-spherical configurations of consistent rotating
black string/ ring solutions and discusses their state-space
geometry, arising from the respective entropy obtained from the
degeneracy of certain bound states, accounted either in terms of
$D$-branes or $M$-branes. Therefore, the study of the equilibrium
state-space geometry characterized by underlying electric-magnetic
charges, dissolved fluxes, with or without angular momenta, is
well suited for an important application of the AdS/CFT
correspondence and determination of vacuum phase-transitions, if
any, in the considered black brane configuration. This is because
the state-space geometry has been one of the central intervention
to explore the thermodynamic structures of large class of black
holes and higher dimensional rotating black brane configurations
\cite{0512035v1,0601119v1,0304015v1,0510139v3,
0606084v1,0508023v2,SST}. See \cite{BNTBull, 0801.4087v1} for
specific instances of various extremal and non-extremal branes,
which have been comprising of the multi-centered branes,
fractional branes, fuzzy rings and bubbling black branes. Ref.
\cite{RotBHs} exclusively has shown that the state-space geometry
spanned by the set of consistent parameters is non-degenerate,
regular and has a negative scalar curvature for the rotating
Myers-Perry black holes, Kaluza-Klein black holes, supersymmetric
$AdS_5$ black holes, holographic corrected $D_1$-$D_5$
configurations and associated spherical horizon BMPV black holes
in five space-time dimensions.

This approach indicates that contemplated configurations are
effectively attractive and stable over a domain of an arbitrary
hyper-surface of their state-space manifolds. It may nevertheless
be noticed that there exists an intriguing relationship between
non-ideal inter-brane statistical interactions and phase
transitions. It may thus be observed that our state-space geometry
analyzes the most probable interactions between the microstates of
the black branes, and typically it determines the nature of
statistical pair correlations, global correlation volume, phase
transitions and certain other reasonable decay, if any, present in
the underlying equilibrium statistical system. These premises may
well be explored from the fundamental duality transformations on
the charges which, in the large compactification radius limit,
imply the existence of general coordinate transformations on an
accompanying equilibrium state-space configuration. An important
ingredient in the present discussion of state-space geometry thus
follows from the fact that there exists a large class of rotating
extremal black branes, non-extremal black branes, multi-centered
black branes and the various other akin black brane solutions in
string theory, $M$-theory and their generalizations which imply a
set of interesting physical configurations. The state-space
geometry is consequently expected to find an intriguing
implication with the AdS/CFT correspondence, and its sound
implication depends on an adequate account of the parameters
defining the degeneracy of a large number of brane microstates
associated with the rotating black strings/ rings.

This article has thus been organized into the following sections
and their respective subsections depending upon the nature of
underlying statistical fluctuations defining state-space
configurations. The very first section provides concerned physical
motivations to study the statistical fluctuations and thereby
initiates why to study the concerned state-space geometry. Section
2 is devoted to a description of the origin of the state-space
approach. In section 3, we briefly outline what is the state-space
geometry, and analyze state-space pair correlation functions,
which may be based on the parameters characterizing an ensemble of
equilibrium microstates for the rotating black brane configuration
under consideration. These specializations are substantially
mimicked in \cite{RotBHs,BNTBull,0801.4087v1}, which give
covariant state-space geometric notions to accompanying
statistical fluctuations about an equilibrium configuration. In
section 4, we introduce the state-space geometry for the charged
rotating black branes in string theory and $M$-theory. In
particular, we shall focus our attention on the underlying
state-space geometry of the various black strings and black rings.
In this prominent section, we consider definite interesting limits
of the supersymmetric black strings, and thereby examine the
statistical correlations, which investigate the phase space
stability and related features of underlying black brane
configurations, in the view-points of string theory and
$M$-theory. Here, we shall precisely divulge basic examples of
three and four dimensional state-space manifolds which, as an
intrinsic Riemannian geometry, may be defined from the negative
Hessian of the entropy of black string and black ring
configurations. Section 5 comprises final remarks and certain
concluding notes of the state-space geometry being anticipated for
the higher dimensional black strings/ rings arising from the
string theory and $M$-theory compactifications. We shall then
discuss the statistical implications associated with the boundary
CFT brane microstates constituting the concerned configurations.
Finally, we confront our construction of state-space manifolds
with the other developments, related role of fluctuations in black
brane configurations, and briefly address some open issues for
future investigations.
\section{$\mathcal N=1$ Supergravity Origin}
In the present section we trace back the origin of the state-space
approach, for non-spherical horizon black holes. The consideration
of the state-space geometry may be described on the basis of a
heuristic argument for predestined black rings and black strings,
which exploits the possibility of higher dimensional black holes
of non-spherical horizon topology. Candidly, the configurations of
present interest may be obtained by taking five dimensional
neutral black strings, constructed as the direct product of the
Schwarzschild solution and a line with the horizon geometry $
\textbf{R} \times S^2$. A circular string may thus be obtained, if
one imagines bending this string to form a circle, and so the
topology now renders to be $S^1\times S^2$. It would thus be
interesting to see how such topology changing effects are
percieved from the perspective of our state-space geometry. In
principle, the circular string thus described tends to contract by
decreasing the radius of $S^1$, which follows simply, due to its
tension and gravitational self-attraction. However, we can make
the string to rotate along the $S^1$ and balance these forces
against the centrifugal repulsion. Then we end up with a neutral
rotating black ring, which as mentioned before, is a black hole
with an event horizon of $S^1 \times S^2$ topology.

Ref. \cite{0110260v2} exhibits an explicit solution where such
objects describe five dimensional vacuum general relativity
configurations. This offers not only an example of non spherical
horizon topology, but also turns out to be a counterexample to the
ordinary four dimensional black hole uniqueness theorem.
Originally, an investigation of rotating black ring solutions has
given firm ground for an existence of vacuum black rings having
rotation along the $S^2$ directions but no rotation along the
$S^1$ ring circle \cite{phir,TMY,PF}. In contrast to the foregoing
rotating rings, the concerned authors have however found that the
charged supersymmetric and non-supersymmetric black rings may be
constructed in a more systematic way, and thus their state-space
manifolds as well are an immediate particular of the present
concern. We further wish to furnish the similar computations for
the general black ring/ strings which may be defined as the
$D$-dimensional black hole having $ S^1\times S^{D-3}$ topology of
(a spatial cross-section of) the event horizon. Correspondingly,
we shall analyze the phase-space structures underlying in these
configurations.

Let us now focus our attention on an intriguing limit of the
state-space geometry associated with the well-known supersymmetric
black ring configurations. To be concrete, we would like to
investigate the nature of state-space geometry of the
five-dimensional supersymmetric black holes appearing in a
subsector of $\mathcal N=1$ supergravity solutions. As mentioned
earlier, these solutions are an example of asymptotically flat
supersymmetric black hole of non-spherical horizon topology and
consequently, it turns out that the underlying state-space
geometry may be uniquely specified by its electric-magnetic
charges and two independent angular momenta. The first example of
our interest is simply associated with an asymptotically flat
supersymmetric solution with regular event horizon having $S^1
\times S^2$ topology. This in fact corresponds to the
supersymmetric black ring configurations, whose existence have
been conjectured in \cite{benakraus} and have further been
physically supported by the investigations of \cite{mathur1,
mathur2}.

It turns out that the state-space manifold of connoted
non-spherical horizon solutions possess a richer structure than
the ordinary BMPV spherical black holes, and in particular, it
motivates us to investigate the nature of state-space
configurations \cite{RotBHs} for the general black rings and black
strings in detail. The fact that the supersymmetry requirements
impose no constraint on angular momenta of the ring solutions
illustrates that the state-space geometry thus obtained may be
co-ordinatized by the electric charges and two independent angular
momenta of underlying equilibrium configuration. Moreover, it
turns out that the solutions under consideration have a
non-vanishing magnetic dipole, which may in an akin limit be fixed
by the asymptotic charges, and therefore it is not an independent
parameter of the concerned solutions signaling a lower dimensional
state-space geometry. Although some black rings are believed to be
unstable, it turns out that the supersymmetry requirement ensures
that very solutions should be stable configurations. These notions
may thus be appraised from the very incentive of our construction
of respective state-space manifolds.

The novel feature of our state-space construction for the
supersymmetric black ring solutions is associated with the fact
that the angular momenta may be determined in terms of the dipole
charges, conserved charges and compactification radius parameter
$R_0$. It turns out that the radius $R_0$ may be determined as the
function of $(J_{\psi}- J_{\phi})$ and concerned dipoles. The
removal of causal pathologies, such as closed causal curves
necessitated in \cite{gghpr}, consecutively initiates further
analysis for the ring configurations, i.e. that the ring entropy
must be a real valued function of the ring parameters, and thus
the state-space configuration must be an intrinsic real Riemannian
manifold. Nevertheless, it is known that the space-like section of
the ring horizon implies that the horizon is geometrically a
product of a finite radius circle $S^1$ and a two sphere $S^2$,
and thus the ring entropy may easily be calculated from the
horizon area as a function of the quantized charges and dipoles.
As a matter of course, one may easily obtain the underlying
state-space quantities and their statistical interpretation for
the large charged supersymmetric black string and black ring
configurations.

It is worth to note that the physical implication of the
non-uniqueness theorem for the supersymmetric black rings and
their statistical configurations force upon us that the problem of
finding supersymmetric solutions may effectively be reduced to
specifying appropriate sources as certain harmonic functions on
the base space manifold \cite{BW}. Physically, these sources
describe $M_2$-branes with both worldvolume directions wrapped on
the internal torus, and $M_5$-branes with four worldvolume
directions wrapped on the torus: i.e. one finds the
$M_2$-particles and $M_5$-strings living along the five noncompact
directions. Moreover, the choice of the harmonic sources
corresponds to a circular loop of $M_5$-strings with a constant
density of $M_2$-particles distributed around the $M_5$-string.
Therefore, it turns out that the parameters being involved
directly lead to an appropriate set of co-ordinates on the
state-space geometry which defines Gaussian fluctuations in the
supersymmetric black ring solutions of \cite{BW}.

What follows next shows that the state-space construction arising
from the supertube configurations entails an admirable physical
interpretation in terms of the parameters of underlying UV and IR
CFTs. Specifically, the UV description indicates that the allied
brane dynamics may be determined by the worldvolume theory which
carries the conserved physical charges. In principle, the UV-CFT
can also describe these black branes with a set of given charges
for both the spherical and non-spherical horizon topology as
different phases of the same theory, in which the dipole charges
act as the parameters of the theory under consideration.
Subsequently, one finds that the underlying theory has different
IR flow, which depends upon the phase of UV-CFT. In fact, the
spherical black hole phase renders to be exactly conformal, while
the black rings induce a non-trivial flow to the IR-theory. It is
nevertheless well known that the central charge of the IR theory
is smaller than that of the UV theory \cite{BK2,KL}. This is
because of the fact that the supersymmetry generators in one of
the chiral sector of the $(4,4)$ UV theory are broken along the
flow, and consequently the IR theory has only $(0,4)$
supersymmetry.

On other hand, the microscopic configuration of the black strings
may be examined by considering the $D_1D_5P$ system whose
parameters define state-space co-ordinates and thus the associated
quantities. This connection is firm, due to the fact that the
usual notion of the string theory implies that the microscopic
description of black holes are typically based on the dynamics of
brane configuration which has the same set of charges as that of
the black hole. In this concern, an intriguing duality exists
between the heterotic string and Type IIA superstrings, which
controls the strong coupling dynamics of the heterotic string in
diverse space-time dimensions \cite{StringTheory}. In fact, an
eleven-dimensional supergravity arises as a low energy limit of
the ten-dimensional Type IIA superstring in the strong coupling
dynamics of string theories. It is however worth to note in
general that the black hole thus obtained may or may not have the
ordinary spherical horizon topology. In fact, the black ring
configurations can carry both the conserved charges and dipole
charges: depending on which of the two sets one puts the stress
on, their description is rather different. Interestingly, it turns
out that the state-space geometry may in these descriptions be
explored further, by investigating whether two intrinsic
state-space manifolds are related by some general co-ordinate
transformations. This is because the description based on
conserved charges is the ultraviolet (UV) completion of the dipole
based description, which in turn completely describes the physics
of the string system at the lowest energies, that is the infrared
(IR) description.

Both of these theories may however be described by usual
two-dimensional sigma-models, and in the extremal limit where the
momentum is chiral, the associated counting entropy formula
follows simply from the Cardy formula which allows one to exhibit
an intriguing state-space configuration. Note that the microscopic
entropy of the rotating non-spherical horizon configurations may
easily be obtained by using the central charge for the momentum
available to distribute among the chiral oscillators, and thus the
underlying state-space geometry would in principle differ in each
description, even if the same object is referred. As mentioned
before, it is the two-dimensional sigma-model, being regarded as
an effective string, whose descriptions differ in what this string
has been taken to be. It is interesting to notice that the IR
theory has an effective string extending along the $S^1$ direction
of ring, and thus the ring may be perceived as a circular string.
The leading order analysis shows in particular that the invariant
state-space global correlation scales as the inverse of the large
charge counting entropy. For the extremal non-supersymmetric black
rings of \cite{RE} and supersymmetric black rings of
\cite{BK2,CGMS}, one finds that there exists an impressive match
between the Bekenstein Hawking entropy and the corresponding
statistical entropy which allows us to give a nice interpretation
of the state-space geometry of typical five dimensional black
brane configurations.

In order to investigate the possible state-space geometric
perspective from the framework of UV theory \cite{CGMS}, we may
further consider an effective string which extends along the
orthogonal dimension to the ring, and thus it follows that the
concerned string configuration may be viewed as the supertube, or
an excitation of it. Such configurations, which have been
developed in \cite{EE} for two-charge black rings, and in
\cite{BK2} for the three-charge supersymmetric black rings, turn
out to be quite interesting for our intrinsic geometric analysis.
We may thus systematically examine the underlying state-space
manifolds in each description and, in turn, a set of equilibrium
parameters defines the associated virtues emanating from the
state-space implications of the general three-charge higher
dimensional rotating black brane configurations. One believes
however that the pursuit of the space-time geometries may be
capturing generic states of the CFT for chosen black holes and
black rings with non-zero horizon area in leading and/ or
subleading orders, and thus capturing enough of them to account
for the horizon area of the macroscopic configurations. Here, we
will use this result to be a crucial ingredient in the realization
of the state-space fluctuations around equilibrium statistical
configurations, for the higher dimensional rotating black string
and black ring solutions.

As a next step towards this objective, it is instructive to
explore the state-space geometry of black rings and relevant other
implications in the limit of the circular string configurations.
Distinguishably, we find that the view points of the IR theory
ascribe an interesting picture to the underlying statistical
correlation functions and related phase space properties over an
equilibrium. This is because the worldvolume theories of black
branes that carry a set of dipole charges have one of the
worldvolume direction along the ring circle which describes the
case of a circular strings, and thus may solely be based on the IR
description of the underlying black ring configuration. The
state-space geometry arising from the consideration of the
microscopic theory turns out to be interesting, in the straight
string limit of the circular rings of finite compactification
radius. This arises from the fact that none of the proposed IR
theories can distinguish between a ring and a Kaluza-Klein
compactified string. Equivalently, the finite size radius
corrections cancel out in the ring entropy, and thus in their
state-space geometry.

It is thus akin to determine a convenient description of the
state-space geometry for an extremal black ring with a finite
horizon, whose entropy may be obtained in terms of the triple
intersection of $M_5$-branes and a running momentum along the
ring. Moreover, the concerned parameters of the solutions thus
achieved determine the central charge of the corresponding
theories, which describe the associated counting entropy and thus
the state-space geometry of the ring solutions. Following
Maldacena, Strominger and Witten (MSW), it turns out in
characteristic that the central charge of the $(0,4)$ theory may
be given by the number of moduli that parameterize the
deformations of smooth intersections of concerned branes, and in
turn is proportional to the number of branes of each kind
\cite{MSW}. One may thus notice that the low energy dynamics can
be described by a $(0,4)$-supersymmetric $(1+1)$-sigma model at
the intersection of the branes, if the six compact directions of
the internal compactifying space are small in the MSW
consideration \cite{MSW}. Furthermore, one finds from the IR
description of such supergravity solutions in the underlying
sigma-model framework that the present interest of the state-space
geometry would be reasonable only when the volume of the
six-dimensional internal manifold is far away from the respective
value of the central charge of the microscopic theory.

An explication of the covariant state-space geometry for small
black ring configuration is thus immediate. In order to do so, one
may provide a string theory description to the structures required
to accommodate different black objects, with the same conserved
charges as that of a class of black rings under question. It turns
out, once again, that the supertube configurations have the right
topology to be identified as the microscopic description of small
fluctuating circular configurations. Both of these configurations
may simply be divulged as an ordinary intrinsic Riemannian
manifold with non-vanishing global state-space correlations. Our
analysis presumes nothing other than the supersymmetric rotating
black rings with given two charges, and one dipole, and thus owing
to the case of usual $K_3\times S^1$ compactifications. In this
concern, the state-space configurations of two charged small black
holes and related string theory and $M$-theory black brane
solutions \cite{BNTBull, 0801.4087v1} show an akin connection with
the non-spherical horizon black holes, what we shall analyze in
the present paper. Furthermore, it turns out that the state-space
geometry provides various useful informations in understanding the
promising nature of underlying statistical configurations
concerning small black holes, black strings, and black rings, in
general. This may easily be justified from the fact that the
underlying microscopic CFT assigns real finite entropy to the
small fluctuating circular black rings. We may therefore identify
that the state-space geometry describes definite fluctuations in
the considered ensemble, and therefore, it renders an illuminating
covariant representation, for the statistical pair correlation
functions and the correlation volume, for the small black ring
configurations.

We may thus anticipate the general consequences of our state-space
investigations, which we shall illustrate for certain natural
black hole/ string/ ring configurations. An important issue of the
classical dynamical stability for generic black brane
configurations turns out to be interesting in this regard, and as
a result, the supersymmetry condition ensures that the
supersymmetric black rings are stable upon quadratic fluctuations
over the parameters defining the chosen configurations.
Furthermore, one may easily realize that the near-supersymmetric
black ring configurations appear as well to have indisputably
similar conclusions under the Gaussian fluctuation in the
parameters which defines an underlying state-space geometry.
Therefore, some qualitative and semi-quantitative arguments may be
possible to make for the small range of the parameters of general
higher dimensional solutions. In this picture, one may easily
observe their global features from the associated scalar curvature
of the state-space geometry. For example, there exist a black ring
configuration, which branches at the cusp between the thin and fat
black ring, and thereby the endeavored configuration must
physically be an unstable black ring configuration. Similar
arguments exist in the literature as well for the other black
rings. In particular, by throwing any small amount of matter which
adds mass but not angular momentum, we may easily see that there
is no other black ring into which this system can evolve and thus
it must back react violently \cite{0110260v2,GRG,HoMy,EEV}.

Most importantly, we may observe that such unstable branchings of
the rings appear at the critical points of the concerned scalar
curvature of the associated state-space manifolds defined in terms
of the parameters of the ring configurations. Qualitatively, one
expects that thin black rings suffer from the Gregory-Laflamme
instability \cite{GL}, which should grow lumps on the chosen ring,
whose evolution may also be governed by the scalar curvature of
the associated state-space manifold. Moreover, our intrinsic
geometric study of the black rings and non-spherical horizon black
holes exhibits that at least one unstable mode is added precisely
at the state-space scalar curvature singularity, when going from
thin to fat black rings. Thus, ring configurations may suffer from
phase transition(s) and other vacuum instabilities, as well, under
the general circumstances. Such a sort of instabilities which we
have been considering here are indeed known in the literature and
thus provide firm support for their concerned vindication. As a
matter of course, these notions are rather well known for the fat
black rings, i.e. that they should again be unstable
\cite{LoTeAr}. Note that the deductions thus made are further
consistent with the known premise of the radial perturbations
\cite{EEV}, against which all the fat black rings appear to be
less stable than the thin black rings.

This alludes that the vacuum black rings with a single spin can be
stable only if the Gregory-Laflamme instability are switched off
and are far from the thin black rings with a small enough angular
momentum. It may also be envisaged that the conclusions thus
obtained for the non-spherical configurations may be generalized
for the general vacuum black rings with two spins. In consequence,
it is known that these higher spin rings also suffer from the
similar superradiant instabilities with a rotating two-sphere
\cite{odias}. Having provided thus a brief account of higher
dimensional black string/ ring configurations, one may note that
there exists an asymptotic expansion of the exact spectrum of the
degeneracy formula for large charges and angular momenta, which
not only reproduces the entropy of the corresponding black brane
solution in the leading order, but also to first few subleading
orders as an expansion in the inverse power of the charges
\cite{0412287, MOOREICTP}. Interestingly, the spectrum of half-BPS
states in $\mathcal N=4$ supersymmetric string theory has adeptly
been divulged \cite{ASen}, and thus one finds an appropriate
notions of degeneracy among a set of correlated microstates. Given
this correspondence between microscopic spectrum and macroscopic
Bekestein Hawking Wald entropy of a black brane configuration, a
natural question would be to understand a microscopic origin of
the state-space geometric interactions arising from the degeneracy
of the CFT microstates. In particular, what does this
correspondence mean for the non-spherical horizon rotating black
branes? This question is however beyond the scope of the present
interest and thus would be relegated for future analysis.
\section{State-space Geometry}
The present section provides a brief account of the state-space
geometry, which as an intrinsic Riemannian manifold $(M, g)$
describes the nature of statistical fluctuations in the higher
dimensional rotating black strings and rings. A similar analysis
has recently been introduced in \cite{RotBHs} for spherical
horizon black holes, which we shall extend here for the black
string and black ring solutions. We shall focus our attention on
correlation functions and correlation volume of four parameter
non-spherical horizon black branes, and thus study the nature of
underlying statistical fluctuations. Besides many more standout
references, the relevant details may be insinuated to
\cite{Weinhold1,Weinhold2,RuppeinerRMP,RuppeinerA20,RuppeinerPRL,RuppeinerA27,RuppeinerA41}
for basic introduction of thermodynamic state-space geometry;
\cite{0512035v1,0601119v1,
0304015v1,0510139v3,0606084v1,0508023v2} for well-known black
holes in general relativity; \cite{SST} for black holes in string
theory; \cite{BNTBull,0801.4087v1} for general state-space
description of diverse pertinent black branes in string theory and
$M$-theory; and \cite{BNTSBVC} for the associated notions of
chemical correlations and concerned relations with the quark
number susceptibilities in $2$ and $3$-flavor Hot QCD
configurations.

Following \cite{RotBHs}, it turns out in the entropy representation that the
components of the covariant state-space metric tensor may be defined to be
\begin{equation}
g_{ij}:=-\frac{\partial^2 S(\vec{X})}{\partial X^i \partial X^j}
\end{equation}
It follows from our outset how the state-space geometry can be
employed to describe fluctuating black ring/ string
configurations. In order to do so, we shall restrict ourselves to
the co-ordinates which are extensive parameters, and then draw
local and global properties of the fluctuations about an
underlying equilibrium statistical configurations. The variables
$\lbrace X^i \rbrace $ shall be thought of, as a set of asymptotic
parameters carried by the higher dimensional black hole and, in
specific, they define a suitable co-ordinate chart on the
state-space manifold. It is worth to mention that the coordinates
$\vec{X}$ are a finite collection, consisting of the mass $M$,
electric-magnetic charges $(P^i,Q_i)$, angular momenta $\lbrace
J_i \rbrace$ and certain local charges $\lbrace q_i \rbrace$, if
any. Thus, the most general co-ordinate, on an arbitrary intrinsic
state-space manifold $(M, g)$, may be expressed as, $\vec{X}= (M,
P^i,Q_i, J_i, q_i) \in M$.

The case of two and three dimensional state-space manifolds have
extensively been described in \cite{RotBHs}. Here, we wish to
explicitly offer the involved intrinsic geometric quantities for
the four parameter black string/ ring configurations. Furthermore,
we may also exhibit that the similar out-lines hold for a class of
general multi-parameter black brane state-space configurations.
For the black ring/ string configurations, the concerned four
dimensional state-space geometry\footnote{ In this section, we
shall invariably use Euclidean notations for an intrinsic
state-space manifold $(M_4, g)$, in the sense that the coordinates
$X_i$, $X_a$ $ \in M_4$ (or corresponding notations $X^i$, $X^a$)
follow contravariant tensor composition rules.} may parametrically
be defined by three possible parameters, {\it viz.}, $\{X_1, X_2,
X_3 \}$, which shall be thought of, as the brane charges, momentum
charges, or dipole charges; and an angular momentum $J$, which
defines a well chosen sector of the microscopic theory. We may
thus expose that the components of the covariant intrinsic metric
tensor are given as
\begin{eqnarray}
g_{X_1X_1}&=& - \frac{\partial^2 S}{\partial X_1^2} \nonumber \\
g_{X_1X_2}&=& - \frac{\partial^2 S}{\partial X_2 \partial X_1} \nonumber \\
g_{X_1X_3}&=& - \frac{\partial^2 S}{\partial X_3 \partial X_1} \nonumber \\
g_{X_1J}&=& - \frac{\partial^2 S}{\partial X_1 \partial J} \nonumber \\
g_{X_2X_2}&=& - \frac{\partial^2 S}{\partial X_2^2} \nonumber \\
g_{X_2X_3}&=& - \frac{\partial^2 S}{\partial X_3 \partial X_2} \nonumber \\
g_{X_2J}&=& - \frac{\partial^2 S}{\partial X_2 \partial J} \nonumber \\
g_{X_3X_3}&=& - \frac{\partial^2 S}{\partial X_3^2} \nonumber \\
g_{JJ}&=& - \frac{\partial^2 S}{\partial J^2}
\end{eqnarray}
Therefore, it is immediate to perceive that the components of the
state-space metric tensor are related to respective statistical
pair correlation functions, which may be defined in terms of the
parameters describing the dual microscopic conformal field theory
on the boundary. This is because the underlying metric tensor
comprising Gaussian fluctuations in the entropy defines the
state-space manifold of the rotating black brane configurations.
Under this consideration, it is worth to note that the local
stability of a fluctuating statistical configuration requires that
the principle components of the state-space metric tensor, $\{
g_{X_aX_a} \mid X_a= (X_i,J)\}$, which signify a set of heat
capacities of the system, should be positive definite
\begin{eqnarray}
g_{X_iX_i} &>& 0, \ i= \ 1,2,3 \nonumber  \\
g_{JJ} &>& 0
\end{eqnarray}
The intrinsic geometric consideration shows that the state-space
fluctuations between the two of the $\{$ branes,  dipoles, angular
momentum $\}$ may be characterized as between two charges, between
a charge and angular momentum, with pure angular momentum. These
fluctuations are simply state-space second moments over the
Gaussian distribution, which has been out-lined in \cite{RotBHs}.
What has been shown for the spherical horizon black holes, we
shall demonstrate here, i.e. that the similar conclusions hold for
the state-space correlations for the non-spherical horizon black
branes, as well. Specifically, the black strings and black rings
show that the diagonal fluctuations involving a set of state-space
co-ordinates are stable, or they rather weaken faster, and thus
come relatively more swiftly in an equilibrium configuration, than
those involving off diagonal state-space fluctuations.

Ref. \cite{RotBHs} presents a detailed analysis of the relative
state-space pair correlations, for the case of two charged
rotating black brane configurations. As promised in the beginning
of this section, in this paper we shall confront with four
parameter black brane configurations. Thus, the desired three
charged rotating black brane solutions can have the three brane
charges and one rotations or three dipoles and one electric
charge,..., etc. Under the present considerations, we may thus
observe $ \forall i \neq j \neq k \in \lbrace X_1,X_2,X_3 \rbrace
$ and, for an angular momentum $J$, that the relative state-space
pair correlation functions form an intriguing qualification
\begin{eqnarray}
C_{ab;cd}&:=& \{\frac{g_{ii}}{g_{jj}},\frac{g_{ii}}{g_{ij}},\frac{g_{ii}}{g_{iJ}},
\frac{g_{ii}}{g_{jk}},\frac{g_{ii}}{g_{jJ}},\frac{g_{ii}}{g_{JJ}},\frac{g_{ij}}{g_{ii}},
\frac{g_{ij}}{g_{kk}},\frac{g_{ij}}{g_{JJ}},\frac{g_{ij}}{g_{jk}},\frac{g_{ij}}{g_{jJ}},
\nonumber \\ && \frac{g_{ij}}{g_{kJ}},\frac{g_{iJ}}{g_{ii}},\frac{g_{iJ}}{g_{jj}},
\frac{g_{iJ}}{g_{JJ}},\frac{g_{iJ}}{g_{ij}},\frac{g_{iJ}}{g_{jk}},\frac{g_{iJ}}{g_{jJ}},
\frac{g_{JJ}}{g_{ii}},\frac{g_{JJ}}{g_{ij}},\frac{g_{JJ}}{g_{jJ}}\}
\end{eqnarray}
It should be noted in particular that the behavior of the set of
independent relative state-space pair correlation functions is
rather asymmetric, in comparison with the identical brane
statistical pair correlations, or the other allowed relative pair
correlations involving charge-charge and rotation-rotation
components. This may nevertheless be physically understood, by the
fact that the brane-brane interaction imparts relatively more
energy, than that of the either existing state-space interactions
and self-interactions. Moreover, we discover, with a non-trivial
refinement, that the possible ratio of state-space pair
correlations, without considering the respective inverse relative
pair correlation functions, may be defined as
\begin{eqnarray}
\tilde{C}_{ab;cd}&:=& \{\frac{g_{ii}}{g_{jj}},\frac{g_{ii}}{g_{ij}},
\frac{g_{ii}}{g_{iJ}},\frac{g_{ii}}{g_{jk}},\frac{g_{ii}}{g_{jJ}},
\frac{g_{ii}}{g_{JJ}}\frac{g_{ij}}{g_{jk}},\frac{g_{ij}}{g_{jJ}},
\frac{g_{ij}}{g_{kJ}}\frac{g_{iJ}}{g_{JJ}},\frac{g_{iJ}}{g_{jJ}},
\frac{g_{JJ}}{g_{ij}}\}
\end{eqnarray}
Following such a physically attractive classification, we may
non-trivially deduce the nature of four parameter relative
state-space pair correlations, and thus indicate the potential
behavior of underlying black string/ ring fluctuating statistical
configurations. This, as anticipated earlier, suggests whether
equilibrium statistical systems are stable under the Gaussian
fluctuations, or not, if either of the charges, dipole charges and
angular momentum take certain specific values. Furthermore, we may
inspect that there exists another choice involving two brane
charges, one dipole and one angular momentum, which may also be
easily carried out in the out-set of the present consideration.

These notions shall be further explicated in the next section for the specific black string/ ring/
supertube configurations, and we shall show that they pertain positive definite state-space correlations,
if the two state-space co-ordinates take particular values. We shall in general analyze the following
two interrogations:
\begin{enumerate}
\item For what value of the brane charge(s), the state-space (relative) pair correlations vanish?
\item Is there some relation of vanishing $f_{ij} \in \tilde{C}_{ab;cd}$,
with the vanishing entropy condition of a given black brane configuration?
\end{enumerate}
For the state-space configurations of spherical horizon black holes \cite{RotBHs}, we know that the
vanishing of distinct parameter state-space pair correlation functions happens exactly at twice the
value of the vanishing entropy condition. We shall explicate these notions for non-spherical horizon black
brane configurations and show in detail that their state-space geometry enjoys a lower bound, on the
attainable brane charges or on the angular momentum, than the constraint arising from the acquainted
large charged leading order black brane entropy solutions. In the next section, we shall show that the
ratios of non-diagonal components vary as inversely, and in turn they remain comparable for a longer
domain of the parameters defining Gaussian fluctuations in the entropy of specified black branes.

Moreover, we may see, from the positivity of the state-space metric tensor, that it imposes a set
of stability conditions on the underlying statistical configuration, over the Gaussian fluctuations.
It is however important to note that the metric conditions are not sufficient to insure global stability,
for the chosen configuration, and thus one may, under this condition, only accomplish a locally equilibrium
configuration. In fact, it is easy to express, in the present case, that the planar and hyper-planar stabilities
of statistical configurations require that all the off-diagonal fluctuations must vanish. This implies that
the principle minors $\lbrace p_2, p_3 \rbrace $ of the underlying metric element should be strictly positive
definite quantities on the entire state-space $(M_4,g)$. An elementary analysis shows that the principle
minors may be represented as the following determinant of the components of the metric tensor:
\begin{eqnarray}
p_2&:=&  \left \vert\begin{array}{rr}
    g_{11} & g_{12}  \\
    g_{12} & g_{22}  \\
\end{array} \right \vert > 0, \nonumber \\
p_3&:=&  \left \vert\begin{array}{rrr}
    g_{11} & g_{12} & g_{13} \\
    g_{12} & g_{22} & g_{23} \\
    g_{13} & g_{23} & g_{33} \\
\end{array} \right \vert > 0
\end{eqnarray}
An existence of positivity of the state-space volume form imposes a stability condition on the Gaussian
fluctuations, over underlying equilibrium statistical configurations, i.e. that one must have $p_4:= \|g\|>0$.
This naturally requires that the  determinant of the state-space metric tensor must be a positive definite
function of the electric-magnetic charges, fluxes, if any, and angular momentum. A straightforward
computation thus yields that the determinant conditions of the metric tensor may be given as
\begin{eqnarray}
\|g\|&:=&  \left \vert\begin{array}{rrrr}
   g_{11} & g_{12} & g_{13} & g_{1J} \\
   g_{12} & g_{22} & g_{23} & g_{2J} \\
   g_{13} & g_{23} & g_{33} & g_{3J} \\
   g_{1J} & g_{2J} & g_{3J} & g_{JJ}\\
\end{array} \right \vert > 0
\end{eqnarray}
This precisely entails that the  state-space determinant and hyper-determinant of the metric tensor
must be positive definite. Thus, the complete stability condition of a black string/ ring configuration
requires that the principle components of the Gaussian fluctuations should be positive definite and the
other components of the fluctuations must vanish \cite{RuppeinerRMP}. In order to ensure this condition,
we may appraise that all the principle minors of the concerned state-space metric tensor must be a strictly
positive definite quantity for the chosen sector of the system. The global stability condition thus forces us
that the set of simultaneous equations expressed by the principle minors must satisfy positivity, in the
domain of interest of the parameters of black hole configuration. In other words, the principle minors
constraints, \textit{viz.} $\{ p_i > \ 0 \ \forall \ i= 1,2,3,4 \}$, should hold under the consideration.

It is worth to note that the determinant condition $\|g\|> 0 $ implies that there exists a positive definite
volume form on $(M_4,g)$. However, in this case, one finds that the classical thermodynamic fluctuation theory
may not remain viable, for the general black string/ ring configurations, see for further details \cite{RotBHs}.
This is because a black brane cannot come into an equilibrium with any extensive,
infinite environment. Thus, the full Hessian determinant of the black brane entropy $S:= S(X^i)$
turns out to be negative definite, when all the asymptotic parameters are fluctuating, see for a motivation
\cite{RuppeinerRMP,Landsberg}. This is also required, in order to produce a local maximum of the total
entropy of the chosen configuration. Thus, the non-existence of a positive definite volume form on the
intrinsic state-space manifold $(M_4,g)$ puts the full problem beyond the control of standard equilibrium
fluctuation theory, when all the state-space co-ordinates $\lbrace X^i, J \rbrace$ are fluctuating.

Physically, one may justify it, by saying that the brane dynamics ultimately favors extreme
situations, where the black brane configuration may either completely evaporate, or grow without limit.
In fact, the black brane will presumably create whatever particles or branes it likes near its event horizon,
and an equilibrium is thus quite unlikely, until the environment is populated by the particles or branes, in
the same proportion to those which have been created. Nonetheless, in a limited domain of the parameters,
when at least one of the parameters is slowly changing in time, with respect to the remaining ones, obliges a
class of stable configurations under the Gaussian fluctuations. This investigation implies a class
of stable configurations with $n$ number of fluctuating parameters, and the remaining
$(4-n)$ parameters being effectively fixed,
or drifting very slowly out of the equilibrium configuration with the environment, where $1\leq n \leq 4 $.

Furthermore, one may note that the scalar curvature corresponding to this state-space geometry elucidates
the typical feature of Gaussian fluctuations about an equilibrium brane microstates of the desired rotating
string/ ring configurations. Furthermore, we may easily procure that the scalar curvature determines a global
invariant on the four dimensional state-space manifold. In this case, it thus follows that one may explicate the
average nature of underlying microscopic black brane configurations. Such an interesting invariant, which accompanies
the information of global correlation volume of underlying statistical systems, turns out to be the intrinsic state-space
scalar curvature. A detailed analysis shows that the general form of the concerned scalar curvature may be expressed as
\begin{equation}
R= \frac{1}{2 \Vert g \Vert^{3}} N(X^i,J)
\end{equation}
The exact expression for the numerator $N(X^i,J)$ is rather involved and thus we relegate it for the
specific considerations. Consequently, it may be noticed that a systematic examination demonstrates
that the state-space geometry with a higher number of charges and angular momenta are similarly not unlikely
to define. In this concern, we remark that one may easily deduce the statistical correlations and
stability criteria, in terms of the corresponding parameters of an ensemble of microstates, which
characterize a possible elementary statistical basis against the macroscopic black brane configurations.

Thence, one may envisage, from the theory of the Gaussian distribution, that the components of the covariant
state-space metric tensor may always be defined as the negative Hessian matrix of the underlying entropy
with respect to the mass, invariant charges, and angular momenta, as well as possible exotic fluxes,
if any, carried by the black brane configuration. It is noteworthy, in the present case, that the black
rings and black strings, when viewed from out-side, may only be characterized by their mass, angular
momenta, and concerned asymptotic charges. Thus, the detailed internal structure and history of their
formations are rather irrelevant, to an asymptotic observer, within the validity limit of the ``no hair''
property. Such a drastic reduction of complexity is clearly an important characteristic of our
state-space investigation for the black string/ ring configurations, whose spherical horizon
case have been considered in \cite{RotBHs}.

Ruppenier has also interpreted, with the assumption ``that all the statistical degrees of freedom
of a black hole live on the black hole event horizon'', that the scalar curvature signifies the
average number of correlated Planck areas on the event horizon of the black brane \cite{RuppeinerPRD78}.
Essentially, the zero scalar curvature indicates certain bits of information on the event horizon
fluctuating independently of each other, while the divergent scalar curvature signals some phase
transitions indicating highly correlated pixels of the informations. Moreover, Bekenstein has
conjointly introduced an elegant picture for the quantization of the area of the event horizon, being
defined as an integral multiple Planck area \cite{Bekenstein}. The concerned examination thus
signifies that the present analysis has a microscopic resolution that rather involves the framework
of well-celebrated Mathur’s fuzzballs \cite{fuzzball}.

Our geometric perspective thus discloses an appropriate ground, with the statements that the state-space
scalar curvature of interest has not only exiguous microscopic knowledge of the black hole configurations,
but also it has an intriguing intrinsic geometric structure. Specifically, we expose that the configurations
under present analysis are effectively attractive or repulsive in general, while they are stable, only
if at least one of the parameter remains fixed. Finding statistical mechanical models with like behavior
might yield further insight into the microscopic properties of black branes. Thus, the hope is that
our state-space analysis would provide a set of conclusive physical interpretation encoded in the scalar
curvature and related intrinsic geometric invariants, for the black stings/ ring solutions.

Incrementally, there exists a series of interesting circumstances, which support the fact that our study
offers a geometrical approach to analyze the microscopic configuration of an ensemble of microstates,
which describe an equilibrium thermodynamic configuration, and thus one may actually apply it to divulge
important physical and chemical behavior of the string theory/ $M$-theory black brane configurations.
In this paper, we specifically wish to consider a general situation of the higher dimensional non-spherical
horizon black hole configurations, where the concerned state-space geometry determines the statistical pair
correlation functions and correlation volume, and thus it may disclose a certain critical nature of the associated
dual boundary conformal field theory.

Thus, the distinguishing features of microscopic acquisitions, which are well-complied with
an understanding of the associated parameters defining the limiting macroscopic configuration,
may easily be exhibited. In fact, it is known that both the apprised descriptions may further be
elucidated from the very likely application of the AdS/CFT correspondence and LLM geometries \cite{LMM}.
General intrinsic geometric appreciation may be envisaged via the associated degeneracy formula
of the black brane microstates, which we shall relegate to other future investigations. The present
consideration nevertheless finds that the state-space scalar curvatures, as a function of the parameters,
entails the central nature of the stabilities of the considered black strings/ rings, while the corresponding
determinant of the state-space geometry, as analyzed earlier, when showing that the most probable states are far apart,
sheds light upon the notion of the degeneracy of an ensemble of brane microstates constituting novel
equilibrium statistical configurations.
\section{Black Rings and Strings}
The present section considers intrinsic geometric aspects of the rotating higher dimensional black branes
and analyzes associated state-space configurations for the case of charged and dipole charged rotating black
strings/ rings. In order to see this point more clearly, we shall focus our attention on the number of branes,
which define the charges associated with five dimensional black string and black ring configurations. Our study
relies on the $SO(4)$ rotation group in $(4+1)$-dimensions, which contains two mutually commuting $U(1)$ subgroups.
This means that it is not unlikely to have rotations in two independent rotation planes, which thus generate
two independent angular momenta $J_\psi$ and $J_\phi$. As often the case in General Relativity, it would indeed
be convenient here too to work with the adapted state-space coordinates, which are possible parameters of the
space-time solutions under question.

To be concrete, we shall specifically consider the ring space-time extending along an insistent
spatial plane, rotating around the $\psi$ direction. Thus, we may easily describe the non-vanishing
contribution of the angular momentum $J_\psi$ into the concerned state-space geometry.
Nevertheless, our general idea would be to find the state-space properties of higher dimensional
black brane configurations. We begin by constructing a foliation of flat space, in terms of the
equipotential surfaces of the field created by a source term, resembling the black string/ ring
one is seeking. For instance, Ref. \cite{HO1} considers an approach to obtain state-space
coordinates suitable for black hole configurations and those of the similar black strings on a Kaluza-Klein
circle. Instead of considering the equipotential surfaces of a scalar field sourced by a ring, it is
sometime more convenient to work directly with the equipotential surfaces of an associated
two form potential: $B_{\mu \nu}$. It turns out that one may regard the ring as a circular string,
which in an underlying effective field theory description acts as the electric source of the three form
field strength $H= dB$.

For example, it is not difficult to construct the solution associated with two form fields, for the
circular electric source of a ring, having Kaluza-Klein radius $R_{KK }$ and phase between $0$ and
$2\pi$ (by using the familiar methods from classical electrodynamics, see \cite{EMT}). In fact, one
finds, in five space-time dimensions, that the electric-magnetic dual of the field $H$ satisfies
\begin{equation}
\ast H= F= dA,
\end{equation}
where $A$ is the standard one-form gauge potential. The dual of the electric string is in fact a magnetic
monopole, which here is just a circular distribution of the monopoles, whose analysis we shall furnish in
the subsection $3.2$. Note further that the space-time invariant/ asymptotic parameters being adapted
for the two form potential $B$, also facilitate greatly the correct choice of the co-ordinate charts for their
state-space manifolds. We shall thus analyze the state-space configurations arising fom the solutions with
gauge dipoles, which have been of definite interest for the case of supersymmetric black rings. For the
purpose of understanding three-charge space-time geometries dual to the microstates of $D_1$-$D_5$-$P$
CFT, one is essentially not so much interested in the black rings with a regular event horizon itself,
but rather in the zero-entropy limit of the rings, which for simplicity we shall refer to, as the three
charge supertubes.

Motivated from \cite{benakraus,bena}, we shall confront with the
supersymmetric black rings, whose existence have also been based
on thought experiments involving supersymmetric black holes and
bubbled black branes, and show almost everywhere that they find a
non-degenerate state-space geometry, which has a non-vanishing
scalar curvature, and thus corresponds to an interacting
statistical system. The subsequent discovery to classify
supersymmetric solutions of five-dimensional $ \mathcal N=1$
supergravity also facilitates several state-space configurations
involving supersymmetric black rings, charged black rings and
other related black brane solutions, see for instance
\cite{EE,EEF}. Thus, the necessary and sufficient condition for
the solutions being supersymmetric is that they reduce to a class
of simple four-dimensional base space manifolds, such that there
exists a canonical form for specific non-spherical horizon black
brane solutions \cite{gu-re-2,gghpr}. In this case again, the
concerned state-space geometry turns out to be generically
well-defined, and corresponds to interacting statistical
configurations. In order to systematically present our analysis,
we shall relegate the detailed investigations of these solutions
to the final subsection $4.4$.

In order to take a closer look at generic black brane microscopic configurations,
one may consider an intriguing idea of bubbling supertubes and foaming black holes, what
Bena and Warner have appraised in \cite{Bena:2005va}. In particular, they have constructed
smooth three-charge BPS geometries, which resolve space-time singularity of the zero entropy
$U(1) \times U(1)$ invariant black ring configurations. The concerned singularity may be
resolved by geometric transitions, which result into the geometries without branes sources or
singularities, but the daughter solution arises with non-trivial horizon topology. The underlying
geometries are both ground states of the black ring, and non-trivial microstates of the
$D_1$-$D_5$-$P$ system. Thus, one finds the form of the space-time geometries, which result from
the geometric transition of the $N$ zero-entropy bubbled black ring solutions. Furthermore,
coincident general geometries, being parameterized by $6N$ functions, give rise to a very large
number of smooth bound-state three-charge solutions, and thus have $6N$ dimensional state-space
configurations. The generic microstate solutions turn out to be specified by the four-dimensional
hyper-K\"ahler geometry of chosen signature, and contain the foam of non-trivial two-spheres.
In turn, there exist definite conjectures that these geometries account for a significant part
of the entropy of the $D_1$-$D_5$-$P$ black holes, and thus the Mathur's conjecture \cite{fuzzball}
may be reduced to counting certain four-dimensional hyper-K\"ahler manifolds.

A general three-charge supertube solution is however given by six arbitrary functions: four of them determine
the shape of the object, three describe the charge density profiles, but the functional constraint coming from
setting to zero the event horizon area makes the choice appropriate \cite{BW,BenaTD}. The near-tube geometry
is of the form $AdS_3 \times S^2$ and, since the size of the $S^2$ is finite, the space-time curvature is
everywhere low. However, since the AdS$_3$ is periodic around the ring, these solutions in fact have a null
orbifold singularity, whose state-space configurations may elsewhere be likewise analyzed. In order to
obtain smooth, physical space-time and state-space geometries corresponding to supertubes given by six
arbitrary functions, one must learn how to resolve the space-time orbifold singularity. Thus, what are
its consequences for the concerned state-space co-ordinates remains an intriguing question. Nevertheless,
we wish to relegate these issues at this point and would like to furnish them at appropriate receptacles.
The specific state-space configurations shall now systematically be divulged for promising black strings
and black rings, and further details are presented in their respective subsections. Finally, the
section $4$ offers a set of definite conclusions, and section $5$ out-lines encouraging remarks for the future.
\subsection{$D_1D_5P$ Black Strings}
In this subsection, we shall initiate our state-space geometric investigation for the simplest rotating
non-spherical horizon solution and compute the underlying quantities, such as the curvature scalar for the
rotating black string configurations for a given Kaluza–Klein momentum. We shall also require that the
concerned electric-magnetic charges, angular momentum, and dipole charges, if any, of the solution have
a fixed symmetry group, and thereby focus our particular attention on the specific sector of the dual $D_1$-$D_5$-$P$
CFT describing three charge black brane configurations with unit Kaluza-Klein momentum.

Most of the progresses in understanding whether the $D_1$-$D_5$-$P$ microstates are dual to the bulk
solutions have occurred on two apparently distinct fronts. The present paper thus investigates the intended
behavior of underlying thermodynamics and statistical mechanics in the perspective of state-space geometry,
and expresses the concerned quantities in terms of the parameters of bulk/ boundary configurations. In fact,
we shall make clear that the first case may be involved in finding two point state-space correlation functions
defined via the parameters of individual smooth solutions carrying brane charges. One thence may go for
analyzing the global properties of the underlying state-space manifold, which may be obtained in terms of the
parameters of the microscopic CFT \cite{GiustoIP,GiustoID, lunin,GiustoKJ,JejjalaYU}. The detailed construction
of the $D_1$-$D_5$-$P$ configurations nevertheless confirms that some CFT microstates, which are dual to the
bulk geometries, have an interesting intrinsic statistical feature, and in fact it seems that the nature of
the equilibrium configuration may be highlighted from the out-set of the state-space geometry.

We shall secondly be involved in understanding the stability of rotating brane configurations, i.e. a physics
issue behind the statistical existence of these CFT solutions, and thus would be interested in analyzing the simplest
definite string theory and $M$-theory configurations from the perspective of the state-space fluctuations.
It has remarkably been shown in \cite{benakraus} that there exists a very large class of black brane
configurations, with three charges and three dipole charges, which can have an arbitrary shape,
and consequently generalize the two-charge supertube considerations \cite{bmpv}. This is because
the entropy of the $D_1$-$D_5$-system comes from arbitrary shapes of the two-charge
supertubes, and thus it is natural to wonder how the arbitrary shapes of three-charge supertubes
account for the sizable part of counting the entropy of $D_1$-$D_5$-$P$ black holes. As a matter of course,
we shall first investigate what is the nature of underlying state-space manifolds for the $D_1$-$D_5$-$P$
black string configurations, with a given Kaluza-Klein momentum charge, and subsequently explore the
consideration for general momentum excitations.

In what follows next, in order to make tangible contact of our state-space geometry with the microscopic
perspective of rotating black objects, we shall consider an enticing case of black string solution
whose microstates may be defined via the parameters of $D_1$-$D_5$-CFT. It is nevertheless known, for
unit Kaluza-Klein momentum, that the counting entropy of deliberated black string is given by
\begin{equation}
S(Q_1,Q_5,J):= \pi \sqrt{Q_1 Q_5- J^2}
\end{equation}
A simple calculation thence finds that the components of the covariant metric tensor,
defined as the negative Hessian matrix of the string entropy, with respect to its
$D_1,D_5$-brane charges and angular momentum $J$, can be expressed as
\begin{eqnarray}
g_{Q_1Q_1}&=& \frac{\pi}{2} Q_5^2 (Q_1Q_5- J^2)^{-3/2}  \nonumber \\
g_{Q_1Q_5}&=& -\frac{\pi}{2} (Q_1Q_5- 2J^2) (Q_1Q_5- J^2)^{-3/2} \nonumber \\
g_{Q_1 J}&=& -\pi J Q_5 (Q_1Q_5-  J^2)^{-3/2} \nonumber \\
g_{Q_5Q_5}&=& \frac{\pi}{2} Q_1^2 (Q_1Q_5-  J^2)^{-3/2} \nonumber \\
g_{Q_5 J}&=& -\pi J Q_1 (Q_1Q_5- J^2)^{-3/2} \nonumber \\
g_{J J}&=& 2 \pi Q_1 Q_5 (Q_1Q_5- J^2)^{-3/2}
\end{eqnarray}
Physically, the principle components of the state-space metric tensor signify heat capacities
or the relevant compressibilities, whose positivity connotes that the underlying system is
in locally stable equilibrium configurations of the $D_1$ and $D_5$-brane microstates.
We may thus observe that the state-space geometry, materializing from the Bekenstein-Hawking
entropy of the $D_1$-$D_5$ black string configurations, admits remarkably simple expressions
in terms of the brane charges and angular momentum. It may thence be suggested that the plausible
microscopic preliminaries might be expected from the state-space configuration, which in the real
sense would be obtained from the Cardy formula or associated general Hardy-Ramanujan formula.
As enumerated in the previous section, one can specifically appreciate the positivity of the principle
components of the present state-space metric tensor, and thus find, for all non-zero brane charges,
\textit{viz.}, $Q_1,Q_5 $ and angular momentum $J$, that they comply
\begin{eqnarray}
g_{Q_1Q_1} &>& 0 \nonumber \\
g_{Q_5Q_5} &>& 0 \nonumber \\
g_{JJ} &>& 0
\end{eqnarray}
%checked
Furthermore, it may easily be seen that the ratios of diagonal components vary as the inverse square of the
underlying parameters which change under the Gaussian fluctuations, whereas the ratios involving off diagonal
components vary only inversely, in the chosen domain of parameters of equilibrium black string configurations.
This suggests that the diagonal components weaken faster and relatively more quickly come into an equilibrium,
than the off diagonal components, which remain comparable for the longer domain of associated parameters
defining the $D_1$-$D_5$ configurations. Importantly, we may easily substantiate, for the distinct
$i, j \in \lbrace 1,5 \rbrace $, that the relative pair correlation functions satisfy
\begin{eqnarray}
\frac{g_{Q_iQ_i}}{g_{Q_jQ_j}}&=& (\frac{Q_j}{Q_i})^2 \nonumber \\
\frac{g_{Q_iQ_i}}{g_{JJ}}&=& \frac{Q_j}{4Q_i} \nonumber \\
\frac{g_{Q_iJ}}{g_{Q_jJ}}&=& \frac{Q_j}{Q_i} \nonumber \\
\frac{g_{Q_iQ_i}}{g_{Q_iJ}}&=& -\frac{Q_j}{2J} \nonumber \\
\frac{g_{Q_iJ}}{g_{JJ}}&=& -\frac{J}{2Q_i} \nonumber \\
\frac{g_{Q_iQ_i}}{g_{Q_jJ}}&=& -\frac{Q_j^2}{2JQ_i}
\end{eqnarray}
%checked
Moreover, it is intriguing to notice that the behavior of the brane-brane statistical pair correlation
function, which is defined as $g_{Q_1Q_5}$ is rather asymmetric in contrast to the other existing pair
correlation functions. In fact, one may understand it, by arguing that the brane-brane interactions
impart more energy than the other captivated  state-space interactions, \textit{viz.}, either
corresponding to the self-interactions or that of the rotations. The present analysis thus proclaims
that the relative pair correlation between the $D_1$-$D_5$ branes, with respect to rotation-rotation
correlations, can be expressed as
\begin{eqnarray}
\frac{g_{Q_1Q_5}}{g_{JJ}}&=& -\frac{1}{4}\bigg(1- \frac{2J^2}{Q_1 Q_5} \bigg)
\end{eqnarray}
%checked
This implies that the relative brane-brane statistical pair
correlation function vanishes exactly at the half value of the
angular momentum, compared to that of the vanishing black string
entropy condition. It may further be explicitly seen, from the
non-vanishing central charge, that the brane-brane statistical
pair correlation is stable, only if the underlying angular
momentum satisfies a lower bound $\vert J\vert< \sqrt{\frac{Q_1
Q_5}{2}}$. Similarly, one may contemplate the nature of other
relative brane-brane statistical pair correlation functions, as
well, and in turn, we find that the ratios, such as
$g_{Q_iQ_j}/g_{Q_iQ_i}$ and $g_{Q_iQ_j}/g_{Q_iJ}$ may easily be
analyzed for all $i,j \in \{Q_1,Q_5\}$.

For to investigate the global properties of fluctuating $D_1$-$D_5$ configurations, we need to determine the stability
along each intrinsic direction, each intrinsic plane, as well as on the full intrinsic state-space manifold.
Intimately, to determine whether the underlying configuration is (locally) stable on state-space planes, one may
thence compute the corresponding principle minors of the negative Hessian matrix of the black string entropy.
In this case, we may easily appraise that the set of principle minors $p_i$, computed from the above state-space
metric tensor, reduces to
\begin{eqnarray}
p_0&=& 1  \nonumber \\
p_1&=& \frac{\pi}{2} Q_5^2 (Q_1Q_5- J^2)^{-3/2} \nonumber \\
p_2&=& \pi^2 J^2 (Q_1Q_5- J^2)^{-2} %\nonumber \\
%p_4&=& 0
\end{eqnarray}
We thus see that the hyper-planar state-space configurations are stable and consequently
the $D_1$-$D_5$ black string configuration remains stable under the Gaussian fluctuation,
if at least one of the parameter is held fixed. Notice further that the determinant of the metric
tensor ($\|g\|=p_4$) is identically zero for the exemplified black string entropy calculation, which
microscopically holds as an asymptotic expansion of the solution, in the limit of large brane charges.
We may nevertheless see that the constant entropy curve, which defines the degenerate state-space
configuration, may be given as \begin{eqnarray} Q_1 Q_5- J^2= c, \end{eqnarray} where the real
constant $c$ may, for a given entropy $S_0$, be defined as $c:= \frac{S_0^2}{\pi^2}$.

In order to extensively appreciate the present consideration, it would further be instructive to extend
our state-space analysis to the most general three charge rotating configurations. In this concern,
the $D_1$-$D_5$-$P$ black branes exhibit that the state-space manifolds associated with BPS black
string/ ring solutions have two microscopic interpretations: one in terms of the parameters of the
$D_1$-$D_5$-P CFT \cite{BK2}, and another in terms of the parameters of the four dimensional black
hole CFT \cite{BK2,CGMS,Bena:2005ni}. This is because the rings on the first hand are described
via the parameters of the black ring CFT, and thus they should be thought out as those of the ground
states of the BPS black ring microstates, in the same way as the parameters of the solutions
of \cite{lu-ma,LuninIZ} arising from the ground state of five-dimensional three charge black holes,
and the other set of solutions of \cite{Bena:2005ay}, yielding the ground state of the four dimensional
four charge black holes, which are dual to the vacua of $D_1$-$D_5$-P CFT. Nevertheless, the present
analysis remains independent of the CFT vacua, and the concerned intrinsic state-space manifolds
correspond to Gaussian fluctuations over the equilibrium configurations. At this point, an exact
establishment of underlying state-space correlations with the CFT relations is left for a future
investigation.

Based on the microscopic description of supertubes \cite{BK2,lu-ma,LuninIZ}, it may however be
anticipated that our geometric construction would correspond to multiple supertubes, which are
dual to CFT states, with longer effective strings than the solutions that come from only one supertube.
Hence, the maximum correlation length in the set up of our intrinsic state-space manifolds, in the
limit of largest number of brane bubbles, should correspond to an ensemble of CFT states with the
longest  effective strings, which are the ones that give rise to the notion of effective state-space
correlations for the $D_1$-$D_5$-P black brane configurations. It is worth to mention that the
state-space geometry stems naturally from the implementation of the machinery of intrinsic Riemannian
geometric Gaussian fluctuations to the most general rotating black brane configurations. Moreover,
the underlying physics in a chosen microscopic sector seems closely related to that of the other
systems containing the correct tower of black hole/ ring/ supertube ground states.

As mentioned earlier, the BPS black rings have two microscopic interpretations:
one in terms of $D_1$-$D_5$-P CFT \cite{BK2}, and the other in terms of the four dimensional
black hole CFT \cite{BK2,CGMS,Bena:2005ni}. Hence, our constructions should similarly regarded, so
that both the microstates of the $D_1$-$D_5$-$P$ system, as well as those of the four dimensional black
hole CFT, indistinguishably describe the state-space of concerned black rings or generic black branes.
The results thus obtained indicate a number of interesting consequences and promising suggestions
for future work. An important exert would consequently be to start from a given $D_1$-$D_5$-$P$
microscopic description of the black ring solutions, and to explore them in the zero-entropy limit.
In turn, our analysis also indicates that the bound states, when the brane could be determined by a
gas of positive and negative centers, with fluxes threading the many non-trivial two-cycles of the
Gibbons Hawking base, and having no localized brane charges on them, would thence characterize
interesting state-space manifolds. This result adheres an appropriate understanding of the
generic horizon changing black brane configurations.

Interesting features of the state-space covariant objects associated with supersymmetric black
strings/ black rings have been based on the fact that these solutions may uniquely be specified
by their electric charges, dipole charges, if any, and angular momenta, which, in the limit of
vanishing compactification radius, render the same conclusion, i.e. that these solutions reducing to
spherical horizon BMPV black hole ones would in general have non-vanishing state-space pair
correlation functions \cite{RotBHs}. In fact, there are no black ring solutions with $J_\psi = J_\phi$,
and thus the co-ordinate charts of the corresponding state-space geometry, as a collection of the parameters
of concerned spherical horizon black brane configurations, would always be different from those of the
conserved charges of the BMPV black holes.

At this point, it may be observed that the entropy, as a function of the parameters
of black ring solutions, is rather maximized in the zero radius limit \cite{benakraus}, and thus
the geometric objects, as the function of angular momenta for fixed and equal brane charges, \textit{viz.},
$Q=Q_i$, may be obtained from the degeneracy which is not continuous in this domain. As we shall
see in the coming subsections, this observation may easily be justified from the reduction
of ring solutions to the BMPV black holes, whose entropy is always greater than the entropy of
the limiting zero radius black ring configurations. As a consequence, there always exists
a discontinuity, which is due to the change in horizon topology from $S^1 \times S^2$ at positive
radius to $S^3$ at zero Kaluza-Klein compactification radius. Analogous geometric observations
may thus be entertained to the discontinuous increase in their entropy, i.e. that the continuity occurs
for the multi-centre extremal Reissner-Nordstr\"{o}m black holes, when the two sources become
coincident. Thus, the reduction of the dimensions of concerned state-space manifold, which is the
matter of the next subsections, is quite natural.
\subsection{Black Rings as Circular Strings}
In order to describe the state-space geometric origin of black ring solution, we may focus on
the corresponding conserved charges adherinf to $M_2$-branes, which can be obtained by turning
the worldvolume fluxes on. In particular, if we set the worldvolume fluxes off, then the ring carries
only $M_5$ dipoles and angular momentum $J_\psi$, which in the absence of fluxes may be identified
with an effective string momentum that defines the central charge of the corresponding aforementioned
IR theory. Although this configuration, at any finite radius, turns out to be non-supersymmetric, even
if the momentum is chiral, however the extremal limit of the dipole ring solution corresponds to the
MSW black branes \cite{MSW}. The present subsection is thus devoted to analyze an associated
state-space behavior of leading order dipoles black ring/ string configurations.

Intriguingly, the world volume theory of branes carrying definite dipole charges, in which the
world volume direction is along the direction of the ring circle, describes an effective extended
string which may be considered as the circular string, after carrying out the associated Kaluza-Klein
compactification. Such a description of the extremal black ring with definite horizon area may be
obtained in terms of the triple intersection of $M_5$-branes, with momentum running along the ring
direction. In turn, the concerned low energy dynamics may be described by the $(0,4)$-supersymmetric
$(1+1)$-sigma model at the intersection of underlying branes. To be specific, let us focus our attention
on the charges that correspond to the case when the fluxes on the $M_2$-branes are turned off, and only
$M_5$-dipoles fluctuate. In this case, the ring state-space configuration can by divulged by only the
$M_5$-dipole charges $\{ n_1, n_2, n_3 \} $ and angular momentum $J_{\psi}$. It is worth to note, in this
case, that the ring solution, in turn, does not correspond to a supersymmetric configuration: neither in
the limit of finite compactification radius nor in that in the limit of the chiral momentum.

A detailed calculation of the infrared microscopic degeneracy formula of generic black rings,
which correctly describes the entropy of MSW black ring solutions, shows in the straight string
limit that the expression of the associated entropy of the extremal black rings may be given by
\begin{eqnarray}
S(n_1,n_2,n_3,J_{\psi})= 2 \pi \sqrt{n_1 n_2 n_3 J_{\psi}},
\end{eqnarray}
which is nothing more than the known leading order large charge macroscopic entropy of the black
string. The details of the concerned estimation of the black string entropy may be found in \cite{MSW}.
Hence, we can compute the components of state-space covariant metric tensor defined as the negative
Hessian matrix, with respect to the three $M_5$-dipole charges $\{ n_i \} $and angular momentum $J_{\psi}$.
The associate components of the state-space metric tensor for the entropy of circular black strings may thus
simply be given as
\begin{eqnarray}
g_{n_1n_1}&=& \frac{\pi}{2n_1} \sqrt{\frac{n_2n_3J_{\psi}}{n_1}}  \nonumber \\
g_{n_1n_2}&=& -\frac{\pi}{2}\sqrt{\frac{n_3J_{\psi}}{n_1n_2}}  \nonumber \\
g_{n_1n_3}&=& -\frac{\pi}{2} \sqrt{\frac{n_2J_{\psi}}{n_1n_3}}  \nonumber \\
g_{n_1J_{\psi}}&=& -\frac{\pi}{2} \sqrt{\frac{n_2n_3}{n_1J_{\psi}}}  \nonumber \\
g_{n_2n_2}&=& \frac{\pi}{2n_2} \sqrt{\frac{n_1n_3J_{\psi}}{n_2}}  \nonumber \\
g_{n_2n_3}&=& -\frac{\pi}{2} \sqrt{\frac{n_1J_{\psi}}{n_2n_3}}  \nonumber \\
g_{n_2J_{\psi}}&=& -\frac{\pi}{2} \sqrt{\frac{n_1n_3}{n_2J_{\psi}}}  \nonumber \\
g_{n_3n_3}&=& \frac{\pi}{2n_3} \sqrt{\frac{n_1n_2J_{\psi}}{n_3}}  \nonumber \\
g_{n_3J_{\psi}}&=& -\frac{\pi}{2} \sqrt{\frac{n_1n_2}{n_3J_{\psi}}}  \nonumber \\
g_{J_{\psi}J_{\psi}}&=& \frac{\pi}{2J_{\psi}} \sqrt{\frac{n_1n_2n_3}{J_{\psi}}}
\end{eqnarray}
One thus observes that the ascertained statistical pair correlations may in turn be accounted for, by
simple microscopic descriptions, being expressed in terms of the number (or dipole charges) of the
branes and angular momentum connoting an ensemble of microstates of the black string configurations.
Furthermore, it is evident that the principle components of the statistical pair correlations
are positive definite, for all the allowed values of concerned parameters of the black strings.
As a result, we may easily see that the concerned state-space metric constraints are
\begin{eqnarray}
g_{n_1n_1}&>& 0 \ \forall \ (n_1,n_2,n_3, J_{\psi}) \mid n_1 > 0 \nonumber \\
g_{n_2n_2}&>& 0 \ \forall \ (n_1,n_2,n_3, J_{\psi}) \mid n_2 > 0 \nonumber \\
g_{n_3n_3}&>& 0 \ \forall \ (n_1,n_2,n_3, J_{\psi}) \mid n_3 > 0 \nonumber \\
g_{J_{\psi}J_{\psi}}&>& 0 \ \forall \ admisible \ (n_1,n_2,n_3, J_{\psi}) \mid J_{\psi} > 0
\end{eqnarray}
%checked
The principle components of the state-space metric tensor $\lbrace g_{n_i n_i}, g_{J_{\psi}J_{\psi}}\ \vert
 i=1,2,3 \rbrace$ essentially signify a set of definite heat capacities (or the related compressibilities),
whose positivity apprises that the black rings comply an underlying local equilibrium statistical
configuration. It is intriguing to note that the positivity of the component $g_{J_{\psi}J_{\psi}}$
requires that none of the brane charges and angular momentum of the associated $D_1$-$D_5$-$P$ IR CFT (or
the charges in the dual $M$-theory description) should be zero in the IR. This is clearly perceptible
because of the fact that the brane configuration becomes unphysical, for these values of the parameters.

It follows, from the above expressions, that the ratios of the principle components of statistical
pair correlations vary as the inverse square of the asymptotic charges; while those of the off-diagonal
correlations modulate only inversely. Interestingly, we may easily visualize, for the distinct
$i,j,k \in \lbrace 1,2,3 \rbrace $, that the interesting statistical pair correlations, thus described,
are consisting of the following scaling properties:
\begin{eqnarray}
\frac{g_{n_in_i}}{g_{n_jn_j}}&=& (\frac{n_j}{n_i})^2 \nonumber \\
\frac{g_{n_in_i}}{g_{J_{\psi}J_{\psi}}}&=& (\frac{J_{\psi}}{n_i})^2 \nonumber \\
\frac{g_{n_iJ_{\psi}}}{g_{n_jJ_{\psi}}}&=& \frac{n_j}{n_i} \nonumber \\
\frac{g_{n_in_i}}{g_{n_iJ_{\psi}}}&=& -\frac{J_{\psi}}{n_i} \nonumber \\
\frac{g_{n_in_j}}{g_{n_kJ_{\psi}}}&=& \frac{n_kJ_{\psi}}{n_in_j} \nonumber \\
\frac{g_{n_in_j}}{g_{n_{i,j}n_k}}&=& \frac{n_k}{n_{j,i}}\nonumber \\
\frac{g_{n_in_j}}{g_{n_jJ_{\psi}}}&=& \frac{J_{\psi}}{n_i} \nonumber \\
\frac{g_{n_in_j}}{g_{n_kJ_{\psi}}}&=& \frac{n_kJ_{\psi}}{n_in_j} \nonumber \\
\frac{g_{n_in_i}}{g_{n_in_j}}&=& -\frac{n_j}{n_i} \nonumber \\
\frac{g_{n_in_i}}{g_{n_jJ_{\psi}}}&=& -\frac{J_{\psi}}{n_j} \nonumber \\
\frac{g_{n_in_i}}{g_{n_jn_k}}&=& -\frac{n_jn_k}{n_i^2} \nonumber \\
\frac{g_{n_in_i}}{g_{n_jJ_{\psi}}}&=& \frac{n_jJ_{\psi}}{n_i^2}
\end{eqnarray}
%checked
As noticed in the previous configurations, it is not difficult to analyze the local stability for
the circular black strings, as well. Radically, one may easily determine the principle minors
associated with the state-space metric tensor. Thus, we may demand that all the principle minors
must be positive definite. Here, we may adroitly compute the principle minors from the Hessian matrix
of the associated entropy concerning the three charge rotating MSW black strings, and thus circular rings.
After some simple manipulations we discover that the set of local stability criteria on various possible
surfaces and hyper-surfaces of the underlying state-space configuration may respectively be determined
by the following set of equations:
\begin{eqnarray}
p_0&=& 1   \nonumber \\
p_1&=& \frac{\pi}{2n_1} \frac{\sqrt{n_2 n_3 J_{\psi}}}{n_1}  \nonumber \\
p_2&=& 0  \nonumber \\
p_3&=& -\frac{\pi^3}{2}J_{\psi} \frac{\sqrt{J_{\psi}}}{n_1n_2 n_3}  %\nonumber \\
%p_4&=& - \pi^4
\end{eqnarray}
%checked
For all physically admitted values of the concerned asymptotic charges (or brane number)
of black strings/ rings, we may thus easily ascertain that the minor constraint, {\it viz.},
$p_2(n_i, J_{\psi})=0$ exhibits that the two dimensional state-space configurations are not
stable, for any value of the brane numbers and assigned angular momentum. Similarly, the
positivity of $p_1(n_i, J_{\psi})$ for arbitrary number of branes shows that the underlying
configurations are locally stable, because of the line-wise positive definiteness.

While the constraint $p_3(n_i, J_{\psi})>0$ respectively imposes the condition that the system
may never attain stability, for any given positive angular momentum and given positive $n_i$'s.
In particular, these constraints enable us to investigate the possible nature of the state-space
geometric stability, for leading order MSW black branes. We may thus observe that the presence of
planar and hyper-planar instabilities exists for the spinning non-spherical horizon circular black
strings. We expect altogether, in the view points of subleading higher derivative contributions in
the entropy, that the involved system demands for a restriction on the allowed values of the angular momentum.

Furthermore, we find that it is easy to enquire about the complete local stability of the full phase-space
configuration, which may in fact be acclaimed by computing the determinant of the state-space metric tensor.
Nevertheless, it is not difficult to enumerate a compact formula for the determinant of the metric tensor.
For the different possible values of brane numbers, {\it viz.}, $\{n_1,n_2,n_3\}$ and angular momentum
$J_{\psi}$, it may apparently be discovered, from the present intrinsic geometric analysis, that the system
finds following expression for the determinant of the metric tensor:
\begin{eqnarray}
g(n_1, n_2, n_3, J_{\psi})= -\pi^4
\end{eqnarray}
%checked
As the determinant of basic state-space metric tensor is a constant and negative quantity, in the large
charge limit. Thus, it requires a non-vanishing central charge of the corresponding $D_1$-$D_5$-$P$
CFT in the IR limit, when the associated worldvolume notion of the dual $M$-theory MSW configuration
holds \cite{MSW}. Our analysis herewith discovers that there exists a non-degenerate state-space geometry
for the leading extremal MSW configurations. However, it is worth to note that the determinant of
the metric tensor does not take a positive definite form, which thus shows that there is no positive
definite volume form on the concerned state-space manifold $(M_4,g)$ of the spinning MSW black brane
configurations, at the leading order contributions in $M_5$-dipoles. This is also intelligible from the
fact that the responsible equilibrium entropy tends to its maximum value, while the same culmination
may not remain valid on the chosen planes or hyper-planes of the entire state-space manifold. It may
in turn be envisaged, in either the $D_1$-$D_5$-$P$ description or dual $M$-theory description, that the
black strings/ circular rings do not correspond to an intrinsically stable statistical configuration.
Thus, it is very probable that the underlying ensemble of CFT microstates, upon subleading higher derivative
corrections, may smoothly move into the more stable brane configurations.

Finally, in order to elucidate the universal nature of statistical interactions and
the other properties concerning MSW rotating black branes, one needs to determine
definite global geometric invariant quantities on the state-space manifold $(M_4,g)$.
Here, we notice that an indicated simplest invariant may be achieved just by computing the
state-space scalar curvature, which may indeed be obtained in a straightforward fashion,
by applying the formerly explained method of our intrinsic geometry. It turns out that
the state-space configuration of the leading order MSW black strings is entirely simple,
and in particular, an explicit expression for the scalar curvature is obtained to be
\begin{eqnarray}
R(n_1,n_2,n_3,J_{\psi})= \frac{3}{2 \pi \sqrt{n_1 n_2 n_3 J_{\psi}}},
\end{eqnarray}
%checked
It is interesting to note that the function $R(n_i,J_{\psi})$ is nothing more than a regular
function, which can completely be defined in terms of the dipole charges and the angular momentum.
In this case, the state-space curvature scalar thus remains a non-zero finite function of the dipole
charges and angular momentum carried by the circular strings, which physically corresponds to an
interacting statistical configuration. We may further observe that the constant entropy curve
and that of the state-space scalar curvature take the same form
\begin{eqnarray}
n_1 n_2 n_3 J_{\psi}= c_i,
\end{eqnarray}
where the real constants $c_i:= \{ c_S, c_R \}$ take obvious values, which may respectively be easily
determined for an assigned entropy and state-space scalar curvature. A simple inspection finds, for
given entropy $S_0$ and state-space scalar curvature $R_0$, that the constants $c_i$ are defined to be
\begin{eqnarray}
c_S&=& \frac{S_0^2}{4\pi^2}  \nonumber \\
c_R&=& \frac{9}{4\pi^2 R_0^2}
\end{eqnarray}
%checked
It is worth to mention that the model may also correctly capture the leading corrections,
which include the effects of the self-interaction among different points along the ring,
that are an effect of the finite ring radius. Thus, the black ring entropy as an
expansion of the inverse compactification radius, although not being an independent
parameter, may rather be fixed in terms of an explicit combination of the large charges.
We thence note that the supersymmetric black rings provide a better behaved statistical system,
and in particular, the finite radius effects appear to be absent in underlying state-space
configurations of the macroscopic entropy formulae of \cite{CGMS,MSW}.

It is thus possible to expose that the black hole non-uniqueness theorems imply a set of important
consequences towards an understanding of the state-space description, and thus it is worth to point
out the nature of Gaussian fluctuations in the MSW ring/ string statistical configurations.
In fact, it is a striking, and perhaps a more important deficiency, inherent to the CFT description,
that the IR-theory does not allow to say anything about the microscopic significance of black
hole non-uniqueness \cite{MSW}, i.e., there are no CFT arguments, how to extract an appropriate set
of parameters. Thus, we may find an ambiguity, when dealing with the intrinsic state-space
manifold of non-spherical horizon black brane configurations.

Interestingly, we find that the determinant of the metric tensor is non vanishing and thus observe
that the underlying state-space geometry describes a non-degenerate intrinsic Riemannian manifold.
It may in turn be carried forward to obtain the resulting scalar curvature arising from the degeneracy
formula, which correctly describes the maximum correlation volume of the black ring configuration, in the
straight string limit. In fact, our analysis easily obtains that the scalar curvature, as a completely
regular function of the dipole charges and the angular momentum, turns out to be inversely proportional
to the entropy. However, it is expected that, to the leading order in $\alpha^{\prime}$, the invariant
quantities would keep a well-defined sign, yielding still a positive definite intrinsic manifold black
strings/ ring state-space configuration.

This means, in a sense, that the dipole-based IR theory looks too closely at the ring configuration,
and by focusing on the string-like aspects of the ring, it is expected that the state-space
configuration is different, with respect to that of the spherical black holes. In order to view both
black objects from a unified perspective, we need to further investigate and observe them from
greater distances, where one may apply AdS/CFT duality to their concerned ultraviolet CFT descriptions
and thus be able to determine an appropriate set of charges/ dipoles and angular momenta, which disport
a dominant role in their weakly interacting state-space configurations.

More generally, in order to saturate the BPS bound, supersymmetric black rings necessarily carry conserved
$M_2$ charges, with integer brane numbers; this, in the microscopic picture, just requires to turn
on fluxes on the worldvolume of the constituent branes. These fluxes, though, give rise to zero
momentum modes that contribute to the total ring momentum, which balances the momentum available
to non-zero-mode oscillators, only if $q_0=-J_\psi$, see for further details \cite{CGMS,MSW}.
This, thus, suggests an intriguing picture for underlying state-space quantities arising from
the microscopic entropy obtained from the Cardy formula and objects compared to the Bekenstein-Hawking
entropy, and thus yields the underlying nature of inclusive $M_2$-$M_5$ limiting configurations.
\subsection{Small Black Rings}
In this subsection, we shall consider the state-space configuration for four parameter small black
rings and focus our attention to analyze concerned state-space pair correlation functions in detail.
In order to provide an appropriate structure required to accommodate state-space quantities
of different black objects with the same conserved charges as those of a class of black rings,
it turns out that the supertube, whose state-space configuration shall be analyzed
in the next subsection, have the right topology to be identified as the string theory description
of limiting small fluctuating circular ring configurations, whose $M_5$-contributions we have
appraised in the previous subsection.

The statistical analysis of  small black rings assumes nothing than the supersymmetric ring, with given
two $D$-brane charges $N_1$, $N_2$, one dipole charge $n_3$, and an angular momentum $J$ for the case
of standard $K_3\times S^1$ compactification and, thence, it turns out that the leading order entropy gives
a precise meaning to our state-space considerations. Moreover, Sen et.al. have shown \cite{0611166} that
there exists an underlying microscopic CFT which assigns the same finite microscopic entropy to the small
fluctuating circular black ring configurations. Thus, it turns out, in both the microscopic and the macroscopic
perspective \cite{MSW,0611166}, that the entropy of a small black ring respectively depends only on the brane
numbers (or related charges) and an angular momentum, \textit{viz.}, $ \lbrace N_1,N_2,n_3,J \rbrace $.
It turns out that the entropy, as a function of the connoted parameters, may be expressed as
\begin{eqnarray}
S(N_1, N_2, n_3, J):= 4 \pi \sqrt{N_1 N_2- n_3 J}
\end{eqnarray}
It is again not difficult to explore the state-space geometry of the equilibrium microstates of the three
charge rotating small black ring, arising from the entropy expression which concerns just the Einstein Action.
As stated earlier, the Ruppeiner metric on the state-space manifold is given by the negative Hessian
matrix of the ring entropy, with respect to the thermodynamic variables. These variables, in this case, are
the rotation and conserved brane numbers, which in turn are proportional to the charges carried by the small
black ring. Explicitly, we may easily find that the components of the covariant metric tensor are
\begin{eqnarray}
g_{N_1N_1}&=&  \pi N_2^2 (N_1N_2-n_3J)^{-3/2} \nonumber \\
g_{N_1N_2}&=&  -\pi (N_1N_2- 2n_3J) (N_1N_2-n_3J)^{-3/2} \nonumber \\
g_{N_1n_3}&=&  -\pi N_2J (N_1N_2-n_3J)^{-3/2} \nonumber \\
g_{N_1J}&=&  -\pi N_2n_3 (N_1N_2-n_3J)^{-3/2} \nonumber \\
g_{N_2N_2}&=&  \pi N_1^2 (N_1N_2-n_3J)^{-3/2} \nonumber \\
g_{N_2n_3}&=&  -\pi N_1J (N_1N_2-n_3J)^{-3/2} \nonumber \\
g_{N_2J}&=&  -\pi N_1n_3 (N_1N_2-n_3J)^{-3/2} \nonumber \\
g_{n_3n_3}&=&  \pi J^2 (N_1N_2-n_3J)^{-3/2} \nonumber \\
g_{n_3J}&=&  \pi (2N_1N_2- n_3J) (N_1N_2-n_3J)^{-3/2} \nonumber \\
g_{JJ}&=&  \pi n_3^2 (N_1N_2-n_3J)^{-3/2}
\end{eqnarray}
%checked
This framework thus affirms that there exists an intriguing intrinsic geometric enumeration,
which describes the possible nature of statistical pair correlations. The concerned state-space pair
fluctuations are determined, in terms of the charges and angular momentum of the dipole black rings.
Hitherto, the principle components of the statistical pair correlations apparently remain positive
definite quantities, for all admissible values of underlying configuration parameters of the black branes.
It may easily be observed that the following state-space metric constraints are satisfied:
\begin{eqnarray}
g_{N_1N_1}&>& 0 \ \forall \ (N_1,N_2,n_3, J) \nonumber \\
g_{N_2N_2}&>& 0 \ \forall \ (N_1,N_2,n_3, J) \nonumber \\
g_{n_3n_3}&>& 0 \ \forall \ (N_1,N_2,n_3, J) \nonumber \\
g_{JJ}&>& 0 \ \forall \ (N_1,N_2,n_3, J)
\end{eqnarray}
%checked
Physically, we may notice that the principle components of the state-space metric tensor
$\lbrace g_{ii}, g_{JJ} \ \vert \ i=N_1,N_2,n_3 \rbrace$ signify a set of heat capacities (or
the associated compressibilities), whose positivity exhibits that the underlying black ring system
is in a local equilibrium statistical configuration of the branes. Our analysis further complies
that the positivity of $g_{JJ}$ obliges that the associated dual conformal field theory living on
the boundary must prevail a non vanishing value of the dipole charge associated with large integers
$N_1, N_2, n_3$, which define the degeneracy of the microscopic conformal field theory.

Interestingly, it is worth to note that our geometric expressions, arising from the entropy of small
black rings, indicate that none of the brane charges can be safely turned off, say $N_i= 0$,
while having a well-defined state-space geometry. However, it is unfeasible to have an intrinsic
black ring state-space configuration with non vanishing charges, say $N_1, N_2 \neq 0$ and $n_3= 0$,
since the objects inside the square-root of the statistical correlations do not take account of the rotation,
and thus argued ring configurations with vanishing brane numbers are no more well-defined black rings,
but they become spherical horizon small black holes, whose state-state is analyzed in \cite{SST}.

The ratio of the principle components of statistical pair correlations form two different sets
and, specifically, they vary as an inverse square of the two involved parameters; while those of the
other off diagonal correlations only vary inversely. It is fundamentally not difficult to inspect,
for non-identical $i,j \in \lbrace 1,2 \rbrace $, that the statistical pair correlations are
consisting of the following type of scaling relations:
\begin{eqnarray}
\frac{g_{N_iN_i}}{g_{N_jN_j}}&=& (\frac{N_j}{N_i})^2 \nonumber \\
\frac{g_{N_iN_i}}{g_{n_3n_3}}&=& (\frac{N_j}{J})^2 \nonumber \\
\frac{g_{N_iN_i}}{g_{JJ}}&=& (\frac{N_j}{n_3})^2 \nonumber \\
\frac{g_{n_3n_3}}{g_{JJ}}&=& (\frac{J}{n_3})^2  \nonumber \\
\frac{g_{N_iJ}}{g_{N_jJ}}&=& \frac{N_j}{N_i} \nonumber \\
\frac{g_{N_iN_i}}{g_{N_in_3}}&=& -\frac{N_j}{J} \nonumber \\
\frac{g_{N_iN_i}}{g_{N_iJ}}&=& -\frac{N_j}{n_3} \nonumber \\
\frac{g_{N_iN_j}}{g_{N_iN_i}}&=& \frac{1}{N_j}(2n_3J-N_1N_2) \nonumber \\
\frac{g_{n_3J}}{g_{N_iN_i}}&=& \frac{1}{N_j}(2N_1N_2-n_3J) \nonumber \\
\frac{g_{n_3J}}{g_{n_3n_3}}&=& \frac{1}{J}(2N_1N_2-n_3J)
\end{eqnarray}
%checked
The concerned microstate counting investigations strengthen the fact that the angular
momentum $J$ of small black rings is further constrained from below, in term of the brane charges
or dipole charges (or numbers), by the bound: $J\leq \frac{N_1N_2}{n_3}$, which defines the
reality condition of the entropy \cite{MSW,0611166}. Furthermore, it is not hard to find, from
the given entropy expression, that the remaining relative state-space correlations retain similar
scaling properties. Here, our analysis thus demonstrates that the positivity of state-space pair
correlation functions between the non-identical charges $\lbrace N_k, n_3 \ \vert \ k= 1,2 \rbrace$
stipulates a modified upper and lower bound on the associated angular momentum
\begin{eqnarray}
\frac{N_1N_2}{2n_3} \leq J \leq \frac{2N_1N_2}{n_3}
\end{eqnarray}
%checked
This is because the brane-brane pair correlations involve the other remaining charges
of the underlying dipole ring configuration. Thus, one establishes that the bounds, arising
from the stability criteria of state-space pair correlations between the non-identical branes,
lie between one half to twice the reality condition of the small black ring entropy.

Note also that our analysis describes a black ring configuration, which is based on
the degeneracy of fluctuations around a circular shape of the limiting supertube, with less
than the maximal angular momentum, $J< N_1N_2/n_3$. Thus, it corresponds to a definite
supersymmetric black ring, which does not suffer from causal pathologies. This condition
intriguingly forces us that the angular momentum of the small dipole black string solution
should statistically satisfy
\begin{eqnarray}
\frac{N_1N_2}{2n_3} \leq J \leq \frac{N_1N_2}{n_3}
\end{eqnarray}
%checked
Apart from the positivity of principle components of the state-space metric tensor,
one further demands that all associated principle minors should be positive
definite, in order to accomplish local state-space stability. It is nevertheless
not difficult to compute the principle minors of the concerned Hessian matrix of
the entropy associated with spinning small black rings. In fact, one encounters, after
some simple manipulations, that the local stability conditions on the one dimensional
line, two dimensional surfaces and three dimensional hyper-surfaces of the concerned
state-space manifold may respectively be measured by the following expressions:
\begin{eqnarray}
p_1&=&  \pi N_2^2 (N_1N_2-n_3J)^{-3/2} \nonumber \\
p_2&=&  \pi^2 n_3J (N_1N_2-n_3J)^{-2} \nonumber \\
p_3&=&  -4 \pi^3 J^2 (N_1N_2-n_3J)^{-5/2}
\end{eqnarray}
%checked
For all physically allowed values of large integers defining the brane numbers/ charges of the
small black rings, we can stipulate that the minor constraint $p_2>0$ obliges that the domain of the
ascribed angular momentum must respectively be greater than $J>0$, for a given non-zero $n_3$ number
of dipoles, while the constraint $p_3>0$ never gets satisfied, for any real physical angular momentum
and entropy. We may thus proclaim that the angular momentum must be positive real valued, in order to
have a definite state-space surface stability condition. As anticipated earlier, we inspect that the
nature of the state-space geometry of the spinning small dipole ring configuration is such that
there exists planar stability, but its state-space is not stable on three dimensional hyper-planes,
for any value of angular momentum.

In addition, it is likewise evident that the local stability of the full small black ring phase-space
configuration may be determined by computing the determinant of the concerned state-space metric tensor.
Here, we may easily compute a compact formula for the determinant of the metric tensor, and indispensably,
our intrinsic geometric analysis provides that the expression for the determinant of the metric tensor, at
the leading order entropy analysis \cite{MSW}, does not find an intrinsic state-space value, for any desired
brane numbers and angular momentum.

It is, in fact, not difficult to see that the determinant of the metric tensor vanishes identically,
for all finite values of the number of branes carried by the small black ring configurations at
this order. Thus, the corresponding state-space geometry turns out to be an ill-defined degenerate
intrinsic Riemannian manifold. Here, the constant entropy curve is given by
\begin{eqnarray}
N_1N_2-n_3J= c,
\end{eqnarray}
%checked
where $c=\frac{S_0^2}{16\pi^2}$ is a known real constant, for a given entropy$S_0$.
Hence, the state-space geometry based on the large charge equilibrium microstates is
trivial, for the leading order small black ring solutions. Also, the present
entropy is unworthy, towards an appreciation of the global statistical correlation length.
It is worth to mention that the present analysis has been obtained from the brane microstate
counting, in the leading non-vanishing entropy approximation, and thus one may anticipate
that the subleading orders contributions would possibly facilitate us a definite,
non-degenerate intrinsic state-space configuration to the dipole small black rings.
Although the associated configuration has a naked singularity instead of a horizon,
however it describes the Bose-Einstein condensate of $J$ short strings of length $n_3$,
in terms of an underlying CFT description of the supertubes, which thus accounts
for the angular momentum in a thermal ensemble of strings \cite{iishi,dabhetal}.

The intrinsic geometric understanding of the small black ring and their statistical configurations
may further be enhanced by the regularization of the space-time singularity, in the virtues of the underlying
string theory/ $M$-theory. Note that our analysis may in turn take account of known specific
considerations, and desired higher-derivative corrections, as well. It may thus be anticipated
that adequate $\alpha^{\prime}$-corrections would ultimately render to well-defined interacting
state-space configurations, for the small black rings. This is because the concerned
low energy effective action requires nothing more than the framework of standard four dimensional
$\mathcal N=4$ supergravity techniques.

Most importantly, we notice that the present analysis relies on corrected horizon configurations,
when the ring is compactified to four space-time dimensions, by putting it on the Taub-NUT space,
which reproduces the familiar Bekenstein-Hawking-Wald entropy of the small black rings, see for instance
\cite{dabhetal}. In turn, De Boer et. al. have shown the existence of a simple model for the dynamical
appearance of Bose-Einstein condensates, which describe an intriguing set of creation operators for
the short strings of length $n_3$ and assigns a dipole of strength $1/n_3$, see \cite{deboer} for
their detailed proposition. Moreover, they have further shown that the one-point functions of
associated dual CFT operators to underlying small black rings are non-trivial. One thus finds that
both constructions support an intriguing association between each other. Consequently,
the state-space description vindicates physically sound containments of the statistical fluctuations,
in small black ring microscopic configurations.

Note further that, if we would like to take an account of the regularization of the singularity
in string theory, then it turns out to be plausible by higher-derivative corrections. In the
framework of a low energy effective action, it again requires nothing more than the usual
techniques of the four-dimensional $ \mathcal N=4$ supergravity backgrounds. In turn,
the modified state-space configurations may be obtained from the corrected horizon, as well,
when the ring is compactified to four dimensions by putting it on a Taub-NUT space, which
reproduces the Bekenstein-Hawking-Wald configuration of the small black rings \cite{dabhetal}.
Moreover, De Boer et. al. have proposed a simple microscopic model for the dynamical appearance
of the Bose-Einstein condensate that describes the creation operators for $J$ short strings of
length $n_3$, which assigns a dipole of strength $1/n_3$, see the details given in \cite{deboer}.
Basically, it may be interesting to see this proposal, how the one-point functions of boundary
operators, in the underlying CFT dual of the small black ring, are non-trivial, in our out-set of
state-space description.

The state-space configurations of a supertube have, as well, been described by definite values
of the specific parameters associated with standard $D_1D_5P$ CFT. In fact, the present
investigation finds that the supertubes are well-defined state-space geometric objects, which
we have explicitly exposed for the single bubbled supertube configurations, having identical
constituent $D_1$, $D_5$ charges, momentum charge and distinct set of dipoles. Thus, in order to
consider the framework of $D_1D_5P$ CFT, one may conveniently start with the black ring configuration
and then pass to a different U-duality frame, which yields a set of required co-ordinates on the
state-space configuration, and thus the parametric perspective of concerned statistical fluctuations.
In precise, the five-dimensional supersymmetric configurations are best understood by, first, uplifting
them to the six dimension black strings \cite{benakraus}, and then viewing them as an intersection
of $D_1$ and $D_5$-branes, which carry non-zero momentum along their common direction. The state-space
correlations associated with the low energies dynamics of the branes may thence easily be described,
via the parameters of an underlying $(1+1)$ UV-CFT with central charge $c_{UV}=6N_1N_2$, where
$N_1$, $N_2$ are the number of the $D_1$ and $D_5$-branes.

Moreover, the underlying statistical configurations of the supertube solutions may further
be well-understood, in the framework of sigma-model CFTs, at those points in the moduli manifold,
where the target space manifold reduces to a symmetric orbifold of $N_1 N_2$ copies
of an internal four-manifold. In turn, one finds that their supergravity description amounts
to a deformation of the theory, away from the associated strong coupling regime. Note further,
that the UV-CFT contains twisted sectors, as well, at the orbifold point. Accordingly, the
statistical configurations associated with the maximally twisted sector physically correspond
to a long effective string of length $N_1 N_2$ times the length of the compactification
circle that the string wraps. That of the untwisted sector may, similarly, be regarded
as containing only the $N_1 N_2$ number of short effective strings. Apart from the long
and short string configurations, it is worth to notice that there exist also some partially
twisted sectors, as well, in the structures of standard $D_1D_5P$-UV/ IR-CFTs.

One thus finds that the physical properties of underlying state-space pair correlation functions
and correlation length acquire an illuminating description for the black strings and black rings;
and as a matter of course, this offers how the local and global statistical correlations fit into
an appreciation of the parameters of the UV theory. However, there is yet another way of understanding
the state-space correlations, in which the angular momentum carried by the $D_1D_5$ brane configurations
can supply one further useful description. One observes wherefore that each individual effective string
has a definite fermionic ground state, which can in general be polarized, and thus is allowed to carry
an angular momentum $(1/2,1/2)$ of the $SU(2)\times SU(2)$ rotation group describing the supertube
configurations. Furthermore, it is clear, for the ground state configurations in the untwisted sector, that
there is a total of $N_1 N_2$ stipulated short strings, which can carry an angular momentum
\begin{eqnarray}
J_\psi&=& N_1 N_2 \nonumber  \\
J_\phi&=& 0.
\end{eqnarray}
One is therefore interested in describing their intrinsic geometric state-space properties.
One finds from \cite{MT} that the angular momentum is present, even in the absence of momentum
excitations. In this case, the concerned ground state as an arbitrary collection of finitely
many brane microstates, which correspond to a class of supertubes, whose state-space configurations
may be shown to be well-defined and interacting statistical systems over a definite range of the equal
$M_2$-brane charge.

Note further that the space-time realization of the supertubes is typically accomplished in terms
of truly existing tubular configurations of branes, which in the case of single bubbled solutions
turn out to be tubes made of single Kaluza-Klein monopole, \textit{viz.}, $n_3=1$, and thus possess
one lower dimensional state-space manifold. In general, if there are several Kaluza-Klein monopoles
with $n_3>1$, then the associated state-space geometry, which has presently been considered, deals with
the various parameters of tubular configurations of constituent branes and may microscopically be
described in specific sectors of the UV-CFT, whose central charge reduces to the value of $c_{UV}/6n_3$.
The state-space configurations restricted in this sector thus deal with those supertubes which involve
strings of length $n_3$, and an angular momentum
\begin{eqnarray}
J_\psi=\frac{c_{UV}}{6n_3}
\end{eqnarray}
For generic supertube configurations, we refer to \cite{EE} for further details and the construction
of their space-time geometries. In order to provide the string theory description of the state-space
structures required for an accommodation of different black objects with the same conserved charges
as that of the supertubes, it turns out that one is obliged to have the right horizon topology, such that
the concerned black brane solutions may generically be confronted with the constituents black ring
microscopic configurations.

In what follows next, we shall focus on the ring solution that corresponds to taking equal values
of the three charges ($D_1$, $D_5$ and Kaluza-Klein momentum) and three distinct dipoles ($D_1$,
$D_5$ and Kaluza-Klein monopole) of a more general supersymmetric black ring (viewed in higher
dimensions, a black supertube \cite{MT}, or the three charge configurations of \cite{benakraus}).
It is worth to anticipate here that the equal-charge solution turns out to be entirely determined
by its conserved charge $Q$, but the case does not remain the same for the most general supersymmetric
black ring or the black supertubes considered in \cite{EEMR2,ElvangRT,Elvang:2005sa}. Thus, an
understanding of how the microscopic string description of black holes distinguishes between the
different horizon topologies has opened an exciting possibility of studying the question of
state-space correlations for the regular horizon black holes with finite horizon area. The
present case in a supersymmetric, highly controlled setting accounts for the state-space
perspective associated with the counting problem and the microscopic existence of the concerned
supersymmetric black ring (/string) configurations.

This may further be supported by the fact that the dipole charge may be defined as an integral
over the constant coordinates surface outside the $S^2$ horizon of black holes. In fact, a given
fixed electric charge determines the radius of $S^2$ and also the angular momentum of the horizon.
It is worth to mention that this charge is not conserved, except in the limit in which the ring becomes
an infinite black string. However, the general solutions of \cite{EEMR2,ElvangRT,Elvang:2005sa} carry
three independent dipole charges, which are respectively proportional to the number of $D_1$-branes,
$D_5$-branes and Kaluza-Klein monopoles, with the worldvolume direction around the ring circle $S^1$.
But the solution of our interest presented here corresponds to taking equal values for the three dipole
charges, so the dipole charge is proportional to the number of branes on the worldvolume direction
around the ring circle. Moreover, one finds in this case that the black rings, when oxidized to six
dimensions, become a class of black supertubes. One may in turn argue that the space-time regularity of
this solution leads to the fact that the dipole charge is being quantized in units of the Kaluza-Klein
compactification radius. An argument follows from \cite{EEMR2}, i.e. that the dipole charges may be defined
by the number of Kaluza-Klein monopoles making up the tube configurations. The next subsection would
thus be devoted to the specific state-space configuration involving a unique large brane charge
supertubes with three dipoles.
\subsection{Supertubes}
In the present subsection we shall consider state-space manifolds arising from the supertube
configurations and analyze the related geometric properties of the underlying statistical
pair correlation functions and correlation volume for $M_2$/ $M_5$-brane configurations.
We shall start our analysis with the supersymmetric black rings whose existence is based
on illustrious experiments involving three charge supersymmetric black holes and supertubes.
Some of the concerned physical motivations arising from the space-time considerations may
be found in \cite{benakraus,bena}. The subsequent discovery shows that there exists an
intriguing program to classify five-dimensional supersymmetric black hole solutions of
$ \mathcal N=1$ supergravity. This in turn facilitates us with the supersymmetric black ring
solutions, charged black rings and other related configurations with non-uniqueness properties,
see for instance \cite{EE,EEF}. It is worth to note that the necessary and sufficient conditions
for supersymmetry reduce to a class of simple base space four dimensional manifolds, such that
there exists a canonical form for the five dimensional supersymmetric solutions \cite{gu-re-2,gghpr}.

The first supersymmetric black ring solutions have chronologically emerged on the flat base
space \cite{ElvangRT} and their subsequent generalization to $U(1)^n$ supergravities are the
matter of \cite{EEMR2,BW,GauntlettQY}. The supersymmetry in fact implies that the mass is
fixed by saturation of the BPS inequality and three Killing fields generating black strings
with $\textbf{R} \times U(1) \times U(1)$ isometry group, and thus demonstrates the same
configurations as that of the non-supersymmetric black ring solutions. These solutions depend
on seven parameters: dipole charges $d^i$, rescaled conserved charges $Q_i$ and the length
scale $R_{KK}$ corresponding to the radius of the ring, with respect to the base space metric.
It is worth to note in the limit $R_{KK} \rightarrow \infty$, with the charge densities $Q_i/R_{KK}$
fixed, that the associated ring solutions reduce to black string solutions \cite{bena}, which are
essentially the same configurations as that of the corresponding neutral cases.

The study of bubbled space-time geometries and axi-symmetric merger solutions turns out to be interesting
for further investigation, from the view-points of our state-space geometry, whose characterization may
be accomplished, in terms of the parameters describing an ensemble of microstates for the supertubes
\cite{07063786V2}. The authors, Bena et. al. have conjointly shown that the microscopic counting entropy,
associated with the single classical black rings, can be expressed as the function of angular momenta,
asymptotically measured electric charges and the number of $M_5$-branes, which characterizes the microscopic
degeneracy of the bubbled black ring configurations. Note further that the classical embedding radius,
which is measured in the space-time $R^2$-plane, exhibits that the angular momentum $J_L$ satisfies
the following constraint:
\begin{eqnarray}
J_L= (d^1+d^2+d^3)R_{KK}^2,
\end{eqnarray}
see \cite{benakraus,bena} for the related details of space-time causal pathologies. It may thus be
noted that the ring charges and angular momenta of the single bubbled ring solutions can be obtained
from possible configurations of the Gibbon-Hawking base points. To simplify the calculation, it is
however better to introduce appropriate state-space variables that make the relation between the
parameters of bubbling solutions and the supertubes more direct. As anticipated for a single bubbled
ring with an identical set of electric charges $Q:=Q_i$, one then finds a simple expression for the
supertube entropy
\begin{eqnarray}
 S(Q, d^1,d^2, d^3)= \frac{\pi}{Q}d^1d^2d^3 \sqrt{4Q-1}
\end{eqnarray}
We therefore observe, in the classical limit $Q \rightarrow \infty $, that the bubbled ring collapses
to the standard ring, and in turn the entropy vanishes corresponding to such a bubbled configuration.
The state-space geometry may thus again be defined as the negative Hessian matrix of the supertube entropy,
with respect to the asymptotic electric charge and number of $M_5$-dipoles. We see in this case that the
components of the state-space metric tensor are
\begin{eqnarray}
g_{QQ}&=& -2\frac{\pi}{Q^3} d^1d^2d^3 (6Q^2- 6Q+ 1)(4Q-1)^{-3/2}  \nonumber  \\
g_{Qd^1}&=& \frac{\pi}{Q^2} d^2d^3 (2Q- 1)(4Q-1)^{-1/2} \nonumber  \\
g_{Qd^2}&=& \frac{\pi}{Q^2} d^1d^3 (2Q- 1)(4Q-1)^{-1/2} \nonumber  \\
g_{Qd^3}&=& \frac{\pi}{Q^2} d^1d^2 (2Q- 1)(4Q-1)^{-1/2} \nonumber  \\
g_{d^1d^1}&=& 0 \nonumber  \\
g_{d^1d^2}&=& -\frac{\pi}{Q} d^3 (4Q-1)^{1/2} \nonumber  \\
g_{d^1d^3}&=& -\frac{\pi}{Q} d^2 (4Q-1)^{1/2} \nonumber  \\
g_{d^2d^2}&=& 0 \nonumber  \\
g_{d^2d^3}&=& -\frac{\pi}{Q} d^1 (4Q-1)^{1/2} \nonumber  \\
g_{d^3d^3}&=& 0
\end{eqnarray}
% checked
Incidentally, we notice from the simple $D$-brane description that there exists an interesting
state-space interpretation, which covariantly describes various statistical pair correlation formulae,
arising from the corresponding microscopic entropy of the aforementioned supersymmetric (extremal)
black brane configurations. Furthermore, our intrinsic state-space pair correlations turn out to be
in precise accordance with the underlying macroscopic attractor configurations, being disclosed in
the special leading order limit of non-vanishing entropy solutions.

In the entropy representation, we thus see that the Hessian matrix of the entropy illustrates
the nature of possible Gaussian correlations between the set of state-space variables, which in this
case are nothing more than the brane charges, dipoles, and angular momenta, if space-time causality
is violated in general. Substantially, we may articulate for given non-zero values of an identical large
electric charge $Q$ and dipole fluxes $\{d^i \mid i= 1,2,3 \}$, as well, that the $g_{QQ}$ is only the
non-vanishing principle component of the present intrinsic state-space metric tensor. It is thence
not difficult to see, for distinct $ i, j, k \in \lbrace 1,2,3 \rbrace $, that the involved component
$g_{QQ}$ satisfies
\begin{eqnarray}
g_{QQ}&>& 0 \ \forall \ (Q, d^i) \mid \{Q \in (3/2, 5/2); \ \{d^i\} \ have \ alternating \ sign \ set \} \bigcup
\nonumber \\ && \ \{Q \in (0,3/2) \cup (5/2,\infty); \ d^i \ has \ different \ sign \ than \ d^j,\ d^k  \}
\end{eqnarray}
%checked
As mentioned before, one finds that the principle components of the state-space metric tensor signifies
respective heat capacities, or the associated compressibilities, whose positivity indicates that the
underlying statistical systems are in local equilibrium of constituent brane configurations.
Furthermore, we perceive that the other allied diagonal components vanish identically, for all electric
charge and dipole charges, \textit{viz,} the dipole-dipole $g_{d^id^i}= 0$. In contrast, the off-diagonal
components do survive in two different sectors. The first one comes as the charge-dipole statistical
fluctuation, while the second comes as the distinct dipole-dipole fluctuation. We may, in either
sector, see easily that their ratio varies as the inverse of the concerned dipoles $d^i$. We further
observe that the first set of statistical fluctuations vanish for $Q=1/2$, while the second set never
vanish, for all finite electric charge $Q$ and non-zero dipole fluxes $d^i$.

This configuration thus shows that the brane-brane fluctuations are rather stable over a domain of $Q$,
and thus relatively more swiftly come into an equilibrium configuration than those involving the off diagonal
ones, which involve brane-dipole or distinct dipole-dipole fluctuations. We may further observe that the
ratios of either non-diagonal components vary inversely, and in turn they remain comparable for the
longer domain of parameters defining the Gaussian fluctuations, in the entropy of specified single bubbled
supertubes. Characteristically, we may easily inspect, in the present case, i.e. $ \forall i \neq j \neq k
 \in \lbrace 1,2,3 \rbrace $ and for a given $Q$, that the non-vanishing relative off-diagonal statistical pair
correlation functions satisfy the following simple scaling relations:
\begin{eqnarray}
\frac{g_{Qd^i}}{g_{Qd^j}}&=& \frac{d^j}{d^i} \nonumber \\
\frac{g_{d^id^j}}{g_{d^id^k}}&=& \frac{d^k}{d^j} \nonumber \\
\frac{g_{Qd^i}}{g_{d^id^j}}&=& - (\frac{d^j}{Q}) \frac{2Q- 1}{4Q-1} \nonumber \\
\frac{g_{Qd^i}}{g_{d^jd^k}}&=& - (\frac{d^jd^k}{Qd^i}) \frac{2Q- 1}{4Q-1}
\end{eqnarray}
%checked
Moreover, it should be noted that the behavior of non-identical brane-brane dipole pair correlations
$\{g_{d^id^j}\mid \forall i \neq j\}$ is rather more asymmetric, in comparison with the other possible
statistical pair correlation functions. This may simply be understood by the fact that the brane-brane
interactions impart more energy than either self-interactions or the correlations, for
given brane charge $Q$. As a result, we discover that the ratio of brane-brane pair correlations, with
respect to unique brane-dipole or unique dipole-dipole relative pair correlations, respectively, turn out to be
\begin{eqnarray}
\frac{g_{QQ}}{g_{Qd^i}}&=& -(\frac{2d^i}{Q}) \frac{6Q^2- 6Q+ 1}{(2Q- 1)(4Q-1)}\nonumber \\
\frac{g_{QQ}}{g_{d^id^j}}&=& (\frac{2d^id^j}{Q^2}) \frac{6Q^2- 6Q+ 1}{(4Q-1)^2}
\end{eqnarray}
%checked
Thus, we deduce that the relative brane-brane correlations vanish exactly at two distinct values of $Q$,
which ares not the same as that of the vanishing entropy condition, nor that of the determinant condition.
This suggests that the brane-brane statistical correlations remain non-zero and respectively stable under
the Gaussian fluctuations, if either of the dipole charge or all of the dipole charges take negative values,
or the $Q$ satisfies an inequality
\begin{eqnarray}
\frac{\sqrt{3}-1}{2\sqrt{3}}< Q< \frac{\sqrt{3}+1}{2\sqrt{3}};
\end{eqnarray}
%checked
or vice-versa with $Q$ not belonging to the above domain. For future applications, it is worth
specifying that the end-points of this interval $(a,b)$ have been determined as the roots of the
equation $6 Q^2- 6Q+ 1= 0$. Similar notions may further be noticed for the relative brane-dipole
correlations, i.e. that they remain positive, iff the two transverse dipole charges take the same sign, and
they vanish exactly at twice the value of the brane charge, with respect to that of the vanishing entropy condition.
Nevertheless, our state-space geometry enjoys lower and upper bounds on the attainable brane charge
than the constraint arising from the acquainted large charge supertube solutions. Furthermore, it may,
independently of the higher dimensional ring rotations, be inferred from our analysis that the supertube
solutions are more strongly saturated than simply the condition of entropy vanishing of the (small) black
branes, or the decoupling limit in the AdS/CFT.

An investigation of definite global properties of the general bubbled ring (or supertube)
configurations determines a certain stability approximation along each direction, each plane
and each hyper-plane, as well as on the entire intrinsic state-space manifold. Specifically,
we need to determine whether the underlying supertube configuration is locally stable on
state-space planes and hyper-planes, and thus one is required to compute corresponding principle
minors of the negative Hessian matrix of the entropy. In this case, we may easily appraise, for
all physically likely values of brane charge and dipole fluxes, that the possible principle
minors computed from the above state-space metric tensor are given by
\begin{eqnarray}
p_0 &=& 1  \nonumber  \\
p_1 &=& -2 \pi d^1 d^2 d^3 (6 Q^2- 6Q+ 1)(4Q-1)^{-3/2} Q^{-3}  \nonumber  \\
p_2 &=& - (\pi d^2 d^3 (2Q- 1))^2(4Q-1)Q^{-4}  \nonumber  \\
p_3 &=& 4d^1d^2 (\pi d^3)^3(Q- 1)Q^{-4}\sqrt{(4Q-1)}
\end{eqnarray}
%checked
Thus, the minor constraints $p_1,p_2<0$ imply that the supertube configurations under
consideration are not stable over the lines, planes of the state-space, for any positive
value of the dipole charges. Similarly, the constraint $p_3>0$ results in an interpretation
that this configuration is stable over three dimensional hyper-planes of the full intrinsic
state-space manifold.

Importantly, we may however ascertain the nature of the state-space geometry of supertube
systems by saying that the planar and hyper-planar stability exists only in a limited domain of
the brane configurations, which arrive whenever one or all of the $d^i$ pick up negative real
values, or the brane charge $Q$ satisfies $ a< Q< b$ and $Q< 1/4$, respectively, for planar and
hyper-planar stability requirement, or vice-versa, such that the minors $p_1,p_2$ of state-space
metric tensor take values over positive reals.

Alternatively, the linear and planar stabilities require that the given supertube configurations
are scarcely populated, and thus the net brane charges are effectively bounded by the rescaled
conserved charges. Moreover, it is not difficult to investigate the global stability on the full
state-space configuration, which may in fact be easily carried out by computing the determinant
of the state-space metric tensor. In this case, we observe that the determinant of the intrinsic
state-space metric tensor is
\begin{eqnarray}
g= \frac{\pi^4}{Q^6} (d^1d^2d^3)^2(12Q^2- 12Q+ 1)
\end{eqnarray}
%checked
which in turn never vanishes, for any given non-zero brane charges, except for the two finite
extreme values of the ring electric charge, $ Q= \lbrace 3/2, 5/2 \rbrace $. We may further
note that the state-space geometry remains positive definite for almost all allowed values of
the electric charge, except in a narrow band when $Q$ satisfies, $ 3/2< Q< 5/2$. We thus observe
that the underlying state-space geometry of bubbled supertubes is in well compliance and, in turn,
corresponds to a generic non-degenerate intrinsic Riemannian manifold.

The concerned metric tensor thus defines a well-defined, almost everywhere non-degenerate, positive
definite state-space manifold, which may solely be parameterized in terms of the brane charge and
dipole fluxes, \textit{viz.}, $ \lbrace Q, d^i \rbrace $. We find however worth to emphasize that
the determinant of the metric tensor indeed takes a positive definite form, except for a region of a narrow
band. Thus, there is no positive definite volume form in this region of the defined
state-space $(M_4,g)$. It may thence be concluded that the supertubes, when considered as the bound
state of branes in the $D$-brane/ $M$-brane description, do correspond to an intrinsically stable
statistical configuration, over a large domain of $Q$.

In order to examine important global properties in these black
holes phase-space configurations, one is further required to
determine the associated geometric invariants of the underlying
state-space manifold. For the present supertubes, the simplest
invariant turns out to be the state-space scalar curvature, which
may as well be easily computed by using the intrinsic geometric
technology defined, as earlier, as the negative Hessian matrix of
the entropy captured by the $M_2$-brane and $M_5$-dipole
contributions. Explicitly, we discover that the concerned
state-space curvature scalar, for the equal charged single bubbled
supertube configurations, may easily be depicted to be
\begin{eqnarray}
R&=& \frac{3Q}{\pi d^1d^2d^3}\{(4Q-1)^{-1/2}(12Q^2- 12Q+ 1)^{-2} \nonumber  \\
 &&(120Q^4- 240Q^3+ 126Q^2- 18Q+ 1)\}
\end{eqnarray}
%checked
Here, we find that the regular state-space scalar curvature seems
to be comprehensively universal for the given number of parameters
of the bubbled ring configurations. In fact, the concerned
perception turns out to be related with the typical form of the
state-space geometry arising from the negative Hessian matrix of
the duality invariant expression of the interested black brane
entropy.

As a standard interpretation, the state-space scalar curvature
describes the nature of underlying statistical interactions of
possible microscopic brane configurations, which in particular
turn out to be non-zero and well defined, for a large band of the
parameters of the chosen supertube solution. Note also that the
absence of divergences in the scalar curvature indicates that the
present supertube solution is an everywhere statistically stable
system as an intrinsic state-space configuration. Thus, it turns
out that there are no phase transition(s), or critical lines, or
any such commensurable phenomena in the underlying state-space
manifold of the black brane supertube solutions, except for the
determinant vanishing configurations. Furthermore, we may easily
appreciate that the constant entropy curve is a rather standard
curve, which may be given by
\begin{eqnarray}
(\pi d^1d^2d^3)^2(4Q-1)= c Q^2,
\end{eqnarray}
%checked
where $c$ is some real constant for the given value of the entropy
of the supertube. This, in fact, determines that the entropy of
the bubbled black branes defines a non-degenerate embedding, in
the view-points of intrinsic state-space geometry. Moreover, we
may also disclose, in the lieu of the present entropy, that the
curve of constant scalar curvature is given as
\begin{eqnarray}
&& 9Q^2 (120Q^4- 240Q^3+ 126Q^2- 18Q+ 1)^2  \nonumber  \\
&& =\pi^2 k^2 (d^1d^2d^3)^2(4Q-1)(12Q^2- 12Q+ 1)^4
\end{eqnarray}
%checked
Here, the constant $k$ is some real number, which may easily be
fixed by giving the value of the state-space scalar curvature.
Moreover, it is not difficult to enunciate that the quantization/
duality conditions, existing on the electric and dipole charges,
signify a set of general coordinate transformation on the
associated state-space manifold, which may be presented in terms
of the net respective parameters of the constituent $D$-branes or
$M$-branes. We see further that the supertubes or associated ring
configurations correspond to a non-interacting statistical system,
for the set of electric charges given by the roots of the
following quartic equation:
\begin{eqnarray}
120Q^4- 240Q^3+ 126Q^2- 18Q+ 1= 0
\end{eqnarray}
%checked
We find, in general, that the state-space geometry $(M_4,g)$ of
single bubbled supertube configurations remains well-defined, up
to another intrinsic Riemannian manifold, $ \tilde M_4:= M_4
\setminus \lbrace 3/2, 5/2 \rbrace $. Thus, there are no critical
phenomena in the state-space manifold spanned by the supertube
parameters, except for the roots of the determinant vanishing
condition. This, in turn, indicates that the underlying
statistical configuration of the specified supertubes is free from
the vacuum instabilities and associated phase transitions.

We find that the state-space geometry of single bubbled supertube
configurations defines a well-defined, non-degenerate, curved,
intrinsic Riemannian manifold, almost everywhere (except for $
3/2< Q< 5/2$) in the allowed domain of the parameters. The single
supertube state-space configuration thus corresponds to an
intrinsically stable, interacting statistical system. It is worth
to mention that the bubbled black ring configurations, in the
classical limit at which they correspond to the standard classical
ring configurations, collapse. Thus, such limits spoil intrinsic
geometric properties of the state-space manifold akin to the
bubbled ring solutions.

Note further that the corresponding microscopic nature of the
state-space manifold may easily be read off, from the position of
the GH-points and the flux parameters associated with the shape of
the ring blob, under the scaling of the true microstates of the
black ring solution, considered in the framework of the abysses
and closed quivers. In this framework, it seems interesting to
investigate the full bubbled configuration space of the $M$-theory
black brane solutions, see for instance \cite{07063786V2}. Such
cases, for example the doubled bubbled black ring configurations,
may find an interesting state-space picture, whose entropy has in
turn been given by
\begin{eqnarray}
S(Q_i,J_L,d^i)&=& \pi [2d^1d^2Q_1Q_2+ 2d^2d^3Q_2Q_3+ 2d^3d^1Q_3Q_1 \nonumber  \\ &&
- (d^1Q_1)^2- (d^2Q_2)^2- (d^3Q_3)^2-3d^1d^2d^3 \nonumber  \\ &&
- d^1d^2d^3\lbrace 4J_L+ 2(d^1Q_1)+d^2Q_2+d^3Q_3)\rbrace ]^{1/2}
\end{eqnarray}
The details regarding the angular momenta $\lbrace J_i \rbrace$, asymptotic $M_2$-charges
$ Q_i \in \lbrace Q_1,Q_2,Q_3 \rbrace$ and concerned $M_5$-dipoles $ d^i \in \lbrace d^1,d^2,
d^3 \rbrace$ may be found in \cite{07063786V2}, which provides a further understanding of the
present investigation from different angles.

Nevertheless, as outlined above for a single bubbling supertube
configuration, it is easy to check that the state-space geometries
apply in more generality, for higher-dimensional black supertubes.
The same may as well be accomplished, for a large class of black
objects with a more complicated horizon topology. We further
expect that these will again match the naive expectation, both in
the large $Q_i$ limit and in the small $d^i$ limit, that the
qualitative nature of underlying statistical configurations would
presumably have very similar conclusions for the general bubbled
backgrounds, as that of the aforementioned singled (or double)
supertube solutions.

We may in fact anticipate that a simple state-space singularity
resolution mechanism, derived under the present consideration,
introduces a set of bounds on the electric charges procured by the
supertube configurations. In order to avoid such instabilities in
the statistical configurations of bubbled black branes, we must
make sure that the state-space manifold, thus associated, must
remain well-defined and everywhere regular intrinsic Riemannian
manifolds. For the singular state-space manifolds, it is however
easy to check that the correlation length diverges and, as we have
noted earlier, that in the present case of a single bubbled
supertube the state-space scalar curvature as a function of the
brane charges, dipole charges and angular momenta, if any, finds a
definite change of signature as the function of thermodynamic
intrinsic variables.

The underlying expressions of the determinant of the state-space
metric tensor and the scalar curvature would also trivially
generalize for the case of more complicated $U(1)^n$ or $SU(n)$
supergravity theories. The general properties of the state-space
geometric equations thus obtained shall presumably be quite
involved, however they have some rather interesting qualitative
features. The detailed explications of these points are however
left out for future exploration. While we have not been able to
analyze a general supertube configuration in a true sense, we
however suspect that the positivity of the determinant of the
state-space metric functions would directly follow from the bubble
equations, triangle inequalities, and some other simple
constraints on the brane and dipole charges of the considered
supertube solutions.

Moreover, the supertube state-space configurations indicate that
the underlying thermodynamic state-space geometry arising form the
Hessian of the corresponding entropy entails an interesting
statistical interpretation, i.e. that they are just state-space
pair correlation functions, which support a definite definition of
the concerned heat capacities (or associated compressibilities).
In most of the cases, the statistical configurations of supertubes
can be understood from our geometric expression involving a large
number of $M_5$-brane dipoles and $M_2$-brane charges, which in
general make up a rotating supertube, with an angular momentum of
the tube alone. It is worth mentioning further, in the case of
zero-entropy black rings, that the concerned angular momentum may
completely be determined in terms of the compactification radius.
In fact, as we have advertised in the previous subsections, our
conclusions thus match exactly with the bubbling supertube radius
and angular momenta, both in the limit when the dipole charges are
small, and in the limit of large electric (or $M_2$-brane)
charges.

The coordinate transformations on the state-space manifold being
defined in terms of the brane parameters imply that one
supergravity solution exists, with two different brane
interpretations. Furthermore, the same is present in all the other
ring systems, as well, whichever contains certain branes wrapped
on topologically trivial cycles, and thus seems to offer definite
key findings of the state-space geometry, which characterizes the
statistical nature of fluctuating bubbling rings and supertubes.
Hence, our state-space investigations, up to the coordinate
transformations, show that the bubbling solutions are identical to
the naive solutions, which physically happens at distances much
larger then the size of the bubble. Moreover, in the classical
limit when the $M_2$-brane charges $Q \rightarrow \infty$, or in
the limit of small number of $M_5$-branes, one finds that the
investigations thus realized, are further supported by the fact
that the bubble configuration, which is nucleated to resolve the
existing space-time singularity of the three charge supertubes,
become very small, and the resolved solution becomes virtually
indistinguishable from the naive brane solution \cite{llm}. This
confirms the intuition coming from the discussion of geometric
transitions, BPS geometries and similar other fundamental
phenomena.

Finding the supergravity solutions of the three-charge supertubes
of arbitrary shape is quite involved. On other hand, it has
however been shown in \cite{BW} that one can solve underlying
equations of motion for these configurations in a linear fashion
\cite{gu-re-2,gghpr,BW}, and may further reduce the whole problem
of finding three-charge BPS solutions to the standard
electromagnetism in four dimensions. In fact, there exists a
side-effect of the study of three-charge supertubes
\cite{benakraus,bena} and the subsequent discovery
\cite{EEMR2,BW,ElvangRT,GauntlettQY} of BPS black rings, which by
themselves have opened up a new window into the prediction of the
black-hole physics \cite{BK2,KL,CGMS,BenaTD,
Bena:2005ni,Bena:2005ay,Elvang:2005sa,
HorowitzJE,GauntlettWH,Gaiotto:2005xt,MarolfCX}. A future analysis
is thus needed to find the nature of statistical fluctuations
beyond the Gaussian approximation, in order to find general
stability criterions in generic bubbled black brane
configurations. This would also indicate an interesting feature of
predisposed statistical fluctuations, i.e. whether an equilibrium
black brane solution truly remains stable and does not suffer from
(vacuum) instabilities.
\section{Conclusion and Discussion}
The present study considers the intrinsic state-space geometry
arising from the fluctuating non-spherical horizon higher
dimensional rotating black brane solutions, and we have
exemplified our out-set for the string theory and $M$-theory
configurations. It is instructive to note that our state-space
investigations are based on an understanding of the microscopic
entropy of diverse black brane configurations, in which the
present consideration requires the coarse graining phenomenon of a
large number of degenerate CFT microstates defining an equilibrium
statistical system. An appropriate analysis thus finds that the
crucial ingredient in analyzing the state-space manifold of
rotating black brane configurations depends on the parameters
provided by an underlying microscopic conformal field theory. Our
illustration of the state-space geometry of higher dimensional
black branes includes the case of $D_1D_5P$ black branes, circular
black rings, small black rings, and supertube configurations. In
such cases of the non-spherical horizon black strings and black
rings, we have thereby focused our attention on the stability
constraints arising from the analysis of state-space pair
correlation functions and global correlation volume.

The state-space geometry thus described may further be shown to
exhibit the associated properties and in particular, one may
analyze the possible physical nature of the state-space geometric
correlation functions and the correlation volume of concerned
statistical configurations. It is worth to mention that the
components of the state-space metric tensor are related to the two
point statistical correlation functions, which are in general
intertwined with the fluctuating parameters of the associated
boundary conformal field theory. This is because the required
parameters of black brane configurations, which describe the
microstates of dual conformal field theory living on the boundary,
may in principle be determined via an application of the AdS/CFT
correspondence. In this way, our intrinsic geometric formalism,
thus described, deals with an ensemble of degenerate CFT ground
states, which at an amusingly small constant positive temperature
form an equilibrium vacuum configuration, over which we have
defined the Gaussian statistical fluctuations.

It is interesting to note that the quadratic nature of Gaussian
statistical fluctuations, about an equilibrium statistical
configuration, determines the metric tensor of associated
state-space manifolds. In either cases, an explicit computation
shows, over the domain of black brane parameters, that the
principle components of state-space metric tensors are positive,
while the non-identical off-diagonal ones may not be so.
Furthermore, in order to appreciate definite global properties of
the concerned configurations, one is required to determine the
nature of stabilities along each direction, each plane, and each
hyper-plane, as well as on the entire intrinsic state-space
configuration.

Our analysis has demonstrated that the determinant of the metric
tensor may as well be negative definite, for the exemplified
higher dimensional black strings and black ring configurations. It
is however in perfect accordance with the known fact of the
Ruppenier geometry that only the classical fluctuations having a
definite thermal origin deal with the the probability
distribution, which has a positive definite invariant intrinsic
Riemannian metric tensor, over an equilibrium statistical
configuration. In fact, our state-space construction, for the
non-spherical horizon black holes dealing with the parameters of
microscopic CFTs, clearly illustrates that the degeneracy and the
signature of the state-space manifold can be indefinite and
sensitive to the location chosen in the moduli space geometry.

Moreover, the absence of divergences in the scalar curvatures
imply that the considered black string/ ring solutions are
thermodynamically stable, and vacuum phase transitions may be
characterized in these configurations. The present investigation
thus serves as a prelude to the state-space geometry of arbitrary
non-spherical higher dimensional black brane configurations.
Intimately, we have explicated that the examples thus explored
have an interesting set up of intrinsic state-space geometry,
which describes the nature of quadratic fluctuations associated
with the statistical configuration of rotating black branes. A
straightforward contemplation finds in general that it is
trustworthy to categorize these configurations in the following
cases.
\begin{enumerate}
\item the underlying configuration turns out to be everywhere well-defined,
whenever there exists a non-zero state-space determinant.
\item the underlying configuration corresponds to an interacting statistical system,
whenever there exists a non-zero state-space scalar curvature.
\end{enumerate}
The main line of thought which has been followed here was,
firstly, to develop an intrinsic Riemannian geometric state-space
geometry conception to underlying leading order statistical
interactions, existing among various CFT microstates of rotating
black brane configurations in string theory and $M$-theory. The
promising perspective of concerned notions thus arise from the
negative Hessian matrix of the corresponding coarse graining
entropy, defined over an ensemble of a large number of brane
microstates, characterizing the considered rotating non-spherical
black hole configurations. Importantly, we have investigated
whether the associated state-space geometries are non-degenerate
and imply an interacting statistical basis for these
configurations, like the one above for instance. The state-space
configuration of supertubes has intriguingly been described by
well separated Gibbon-Hawking charges, with vanishing total
Gibbon-Hawking charge. We further divulged that the behavior of
statistical pair correlations, among the equilibrium microstates,
may be governed by the consistent parameters defining concerned
microscopic CFT vacuum configurations. Thence, the same remains
valid for the other associated intrinsic geometric quantities, as
well, on the respective underlying state-space manifolds.

In this article, we have explained how particular statistical pair
correlation functions, associated with an ensemble of CFT
microstates, characterize the behavior of state-space manifolds of
the higher dimensional rotating black string and black ring
configurations. Interestingly, it has been admissible to
investigate whether the state-space correlations, as the function
of desired brane numbers or charges on them, are positive definite
and regular/ singular function on the respective state-space
manifolds. Moreover, in the case of a singular state-space scalar
curvature, we have analyzed the nature of concerned singularities,
and thus presented them as critical curves on the underlying
state-space equilibrium manifolds. Thereupon, we have shown that
there exists definite critical points, critical lines, and
critical (hyper)-surfaces, which may all be defined by the
divergence structure of underlying state-space scalar curvatures.

After having a detailed discussion of non-spherical horizon
topology black holes, we have explicated the nature of their
underlying state-space configurations. In this concern, the
preliminaries which have been motivated in \cite{RotBHs} have been
extended for an explication of generic state-space pair
correlation functions, in general. For example, the
$D_1$-$D_5$-$P$ black strings and small black strings indicate
degenerate state-space geometry, while the circular strings and
supertubes have non-definite state-space configuration. It is
nonetheless expected that an appropriate contribution of higher
derivative corrections would facilitate numerous black strings/
ring configurations, having well-defined positive definite
state-space geometry. The higher order
$\alpha^{\prime}$-corrections, when taken into account in the
underlying effective theory, are envisaged to offer diverse
well-defined state-space configurations or, at least, the
non-degeneracy of state-space configurations may generically be
expected, in comparison with the ill-defined state-spaces obtained
for the $D_1$-$D_5$-$P$ black strings and small black strings, at
the leading order entropy solutions. In addition to this, similar
conclusions have been established for the heterotic small black
holes \cite{SST}, and thus we contemplate that the black strins/
rings should acquire a well-defined, non degenerate and curved
state-space geometry. There are however many caveats, many things
which require further clarification and many open questions, and
we leave such issues for the future.

The present article analyzes the case of general higher
dimensional configurations and thereby shows that the black
strings/ rings may not have well-defined state-space manifold, in
general. Some of the arguments are briefly based on the nature of
parameters, which characterize a countably large number of
degenerate equilibrium microstates and thus define concerned
rotating black brane configurations. The specializations made are
analogous to those mimicked in \cite{BNTBull,0801.4087v1}. Here,
we have defined the notion of statistical interactions in the
context of bubbling AdS geometry and other well-known black string
and black ring configurations. In section $2$, we have
specifically developed what are the relative state-space pair
correlation functions for given configuration. We have explicitly
constructed general features of our state-space geometric
covariant quantities, which come from an underlying black brane
entropy, and then discussed them with those that arise from their
naive microscopic investigations.

The state-space manifold thus described primarily out-lines the
stability criteria over the range of parameters defining the
concerned higher dimensional black brane configuration, with or
without rotations. Indeed, we find that the intrinsic geometric
implications arising from the state-space construction are
stunning. In particular, they provide a unified framework to
analyze both the statistical correlations and singularities, which
in general may exist in arbitrary finitely many parameter black
brane configurations. In this case, we notice definite interesting
supergravity configurations, satisfying well-definiteness
condition for the state-space geometry, and thus supplying a
globally well-defined positive definite volume form to the
concerned state-space  manifold. It has explicitly been shown in
the large charge limit, where underlying computations of the
entropy are valid, that the supertube configurations have a
positive definite metric tensor and a specific scalar curvature.
Our exposition thus provides simple enough techniques to disclose
the nature of fluctuating black string and black ring
configurations.

Note that state-space conclusions, in general, depend on the
attractor fixed points of the underlying moduli space
configuration, and thus they can vary over basin to basin, or away
from either attractor fixed point(s) of chosen black brane
attractor solutions. The present analysis nevertheless finds
ground in the leading order contributions, and it turns out that
the obtained certitudes remain valid under the perturbative
approximation of the black brane entropy, in a chosen attractor
basin. It may further be expected that nothing special would
happen, and the number of degenerate CFT states forming the vacuum
does not change, over the domain of the parameters of large black
brane configurations. Another important ingredient in the present
discussion of state-space geometry has been the fact that there
exists a large class of rotating extremal black branes,
non-extremal black branes, multi-centered black branes and various
other tenable black brane configurations in string theory and
$M$-theory, which offer several interesting physical implications.
It may also be envisaged that the state-space geometry may
interestingly find parametric implications with other
developments, and the AdS/CFT correspondence seems perfect, in
this regard, that it may offer an adequate account of the
fluctuations among an ensemble of brane microstates of rotating
black branes and thus their associated state-space configurations.
\subsection*{Black String and Black Ring Intrinsic State-spaces}
The state-space geometry of five dimensional supersymmetric black
ring solutions may in general be described by five independent
conserved charges, \texttt{namely}, the $D_1$, $D_5$ brane charges
and momentum charge (determining the mass through the saturation
of the BPS bound), and two independent angular momenta. In an akin
limit of charges, one can easily make contact with the string
theory or $M$-theory, by arguing that the net charges and dipole
charges must be large quantified integers, since they represent
the number of branes and units of Kaluza-Klein momenta. As the
consequence of charge quantization, we thus observe that the
finite violation of black hole uniqueness theorems, for
non-spherical horizon solutions, would in fact simply enter in
their state-space geometric properties.

Moreover, it may be observed that the present study relies on the
analysis of the entropy of black strings and black rings, which in
the original string theory computations \cite{9601029v2} may
easily be accounted for those microstates, which have the same
conserved charges as that of the concerned black brane
configurations. However, the entropy calculation shows that the
black ring configurations always have an unequal amount of angular
momenta, and thus the rings themselves go on the right track with
the proposal of \cite{EE}, and thus with their state-space
geometric investigations, what we have been analyzing here and
exploring for the future. The concerned features may essentially
be elucidated from the fact that we should focus on the counting
of only those microstates, which belong to the specific sectors of
$D_1$-$D_5$ CFT, and in turn, the precise sector may be determined
by assigning definite values to the dipole charge(s) and angular
momenta.

On other hand, we find that the black brane configurations
describing black rings and black strings indicate similar
statistical pair correlations and intrinsic geometric properties
to those of the spherical horizon black holes \cite{RotBHs}. In
fact, the present analysis is surmised to provide positive
evidence for physically significant identifications associated
with the attractor mechanism, and that of the microstates counting
of the chosen configurations. In short, the present study provides
an introduction to the state-space of black branes in general
relativity, string theory and $M$-theory. The very purpose of
present the inquiry, thus, explicates the typical nature of
state-state configurations, which arrive with the five-dimensional
black holes, of $S^1\times S^2$ event horizon topology. Novel
aspects of our presentation further include an intriguing approach
to construct state-space manifolds, arising from the parameteric
fluctuations of black ring/ black strings, i.e. that an ensemble
of CFT microstates of chosen black branes furnish a critical
reconnaissance of the underlying equilibrium state-space
configuration.

An existence of supersymmetric black ring configurations thus
raises a set of interesting questions, for an investigation of the
intrinsic state-space geometry. For example, the present needs may
worthfully be examined, by considering
\begin{enumerate}
\item what happens to the state-space manifolds, for the supersymmetric black holes of the minimal
supergravity solutions in the empty regions of the $J_\phi$-
$J_\psi$ plane, or any other solutions which overlap with the
covered regions?
\item what are the implications of state-space manifolds, with the full uniqueness theorem being
strengthened by the \cite{reall:02}, for the supersymmetric black
holes and black supertubes?
\end{enumerate}
It would also be important to see whether this kind of approach
can be pushed further to general non-extremal black ring
solutions, which reproduce the above solutions as a special case;
and those of the \cite{0110260v2,HE}, which presumably depend on
the five parameters corresponding to angular momenta,
electric-magnetic charges, dipole charges, in general, and mass of
the black string and black ring configurations under the question.
Finally: is it possible to perform detailed statistical mechanical
or CFT calculations, against the state-space pair correlation
functions and correlation volumes associated with the extremal
black string and black ring solutions? It would further be
instructive to extend our analysis for the most general
non-extremal and/ or non-supersymmetric black brane
configurations. In order to appreciate the present consideration,
what we have focused in this paper may in general be analyzed for
generic black strings, black rings and supertubes. In the next
section, we shall thus offer a set of prospective issues, for the
state-space configurations of general $D_1$-$D_5$-$P$ and
$M_2$-$M_5$ configurations.
\section{Future Directions and Open Issues}
Promising explorations of state-space geometry to finite parameter
black string/ ring configurations in string theory and $M$-theory
offer the definite explicit nature of generic non-spherical
horizon black strings and black rings. Before concluding this
article, we thus wish to address some recent developments in the
string theory and $M$-theory. The rest of this section would focus
its attention on very general interests, possible implications and
analyze some of them, from the perspective of our intrinsic
state-space geometry.
\subsection*{(i) State-space Instabilities and dual CFTs}
Definite progresses have been made towards the construction of
3-charge solutions describing black ring microstates, and in
particular the method of adding a small amount of Kaluza-Klein
momentum, as a perturbation to 2-charge supertubes results in the
3-charge solutions of \cite{3chargepert}. In this paper, we have
investigated underlying parametric local and global statistical
correlations, for the generic black string/ ring configurations
under specific considerations, and thereby constructed their
related state-space geometries. Furthermore, our state-space
description is expected to provide the nature of a large class of
three charge ring solutions, which are non-spherical horizon black
holes, whose microstates have been proposed and studied in the
footings of \cite{lunin,GiustoKJ,Bena:2005va,Berglund:2005vb}. In
fact, such an analysis accounts for the covariant description of
the configurations being considered in Refs.
\cite{Bena:2005va,Berglund:2005vb}. Upon applying aforementioned
techniques, one can explicate the nature of multi-center
Gibbons-Hawking solutions, whose base space manifold deals with
various poles of positive and negative residues.

These solutions describe rich topological structures related to
the dipole sources resolved into fluxes, along new internal cycles
of the moduli configurations. However, the mapping between these
supergravity solutions and their dual CFT states has not really
been properly identified in pertaining cases in the existing
literature. Thus, several akin micoscopic meanings of the
state-space geometric investigations are yet stimulated to be
examined into the particulars of the associated dual CFTs and of
the microscopic duality symmetries. The issues like string duality
symmetries and their macroscopics interplay with our parametric
intrinsic state-space geometry have been left out for the
associated general rotating black brane configurations appearing
under the string theory and $M$-theory considerations. As for the
limiting configurations being divulged in this paper, we may
nevertheless ratify that the pertinent instabilities, if any,
would precisely occur at the divergence points, curves, and
hyper-surfaces of their state-space geometric invariants of the
generic rotating and non-rotating black brane solutions.

Qualifications of dual CFTs and microscopic symmetries deserve
further study and, in turn, they leave many open possibilities,
under which the black brane configurations can acquire various
instabilities in the domain of their parameters. It is worth to
mention, thereupon, that the dipole charges of given black black
brane configurations may help to stabilize them against some local
and/ or global perturbations, like GL modes, chemical potential
fluctuations corresponding to the electric-magnetic charges and
dipole charges, rotational fluctuations corresponding to the
angular momenta, and the thermodynamic temperature fluctuations
for the near-extemal and non-extemal black brane solutions.
Thence, a large class of such affiliated instabilities, connoted
by general rotating black brane configurations in higher
space-time dimensions, deserve further study.
\subsection*{(ii) $D$ Dimensional Black Brane Configurations}
We further see that number of parameters which characterizes a
black brane configuration increases with the space-time dimension
$D$, and thus the degrees of freedom of gravity increase
accordingly. It is therefore natural to expect more complex
dynamics possessing more parameters of the solution, and thus
suggesting a relatively higher dimensional state-space manifold,
when the space-time dimension $D$ grows. It is known that the
standard $D=4$ gravity is highly constrained, due to the black
hole uniqueness theorems, and thus one finds that the only
rightful parameters are the mass, charges, and angular momentum,
which completely parameterize the existing black hole space-time
solutions. It is thus anticipated that one may easily define and
compute concerned invariants of underlying state-space manifolds
and in principle analyze the possible statistical nature of the
known $D=4$ black holes
\cite{RotBHs,BNTBull,0801.4087v1,0606084v1,SST}.

The discovery of black strings/ rings shows however that $D=5$
solutions allow more freedom. Their space-time dynamics is still
amenable to some additional dipole charges, in contrast to that of
the spherical horizon black hole solutions of $D=4$ gravity. In
this paper, we have performed a detailed study of the underlying
state-space manifolds, for various contrivable possible cases of
their parameters, which characterize chosen black brane
configurations and arise as the consistent lower dimensional
solutions, in the view-points of string theory and $M$-theory
compactification(s). It is thus natural to wonder what would be
the story for the $D>5$ gravity configurations, which largely
remain unexplored, although there have been some indications that
black holes (even spherical horizon ones) may possess
qualitatively unexpected features \cite{RotBHs,EM}. The concerned
issues for the general black brane configurations would thus be
the matter of future investigations.

It would indeed be interesting to analyze concerned behaviors of
the associated state-space manifolds, for various black rings of
horizon topology $S^1\times S^{D-3}$, for $D>5$. Moreover, we
would like to explore what happens to them, with some other
horizon topologies in $D>5$. For example, the consistent higher
dimensional horizon topologies, which turn out to be the cases of
$S^1\times S^1 \times S^2$, $S^3\times S^3$, are the open
possibilities, in accordance to the higher-dimensional horizon
topology theorem \cite{topology}. Physically speaking, the
dimension-dependent gravitational force decays faster with the
distance, so it would play a dominant role only around an
equilibrium statistical system for the small radii rings. In turn,
it has been suggested that the thin black rings should thus exist
in $D>5$, as in standard $D=5$ solutions, and they should also be
unstable against radial perturbations \cite{topology,MP}. We hope
here that these notions may incidentally be explained from the
divergence structure of underlying state-space geometries.

Observe further the fact that there is no bound on the angular
momentum, even for the Myers Perry black holes with single spin in
$D>5$ \cite{MP}, and so an existence of higher dimensional
rotating black strings/ rings would automatically imply a definite
violation of the black hole uniqueness properties. This in turn
implies that the underling higher dimensional state-space
manifolds would relatively be parameterized by a large number of
dipole charges, which live on an equal footing with the other
globally conserved parameters of candid holes, rings, tubes,...
and other pertinent configurations. Thus, there indeed exists the
definite possibility of an even larger class of instabilities in
the higher dimensional black brane configurations, which need to
be understood in the framework of our state-space geometry.
\subsection*{(iii) Bubbling Black Brane Solutions}
The physics of space-time singularity resolution is very
preeminent to the present state-space motivation. In fact, what
has been inspired by the investigations of Lin, Lunin and
Maldacena (LLM) \cite{llm} finds a sound physical consideration in
which the bubbling solutions reduce to naive giant gravitons in
the small dipole charge limit, and thus have topological
transitions where constituent branes are replaced by dissolved
fluxes. It is important to note that the indicated black branes,
having well separated Gibbon-Hawking charges and vanishing total
Gibbon-Hawking charge, have non-degenerate and curved state-space
configurations, whose intrinsic metric tensor, as a result,
determines heat capacities. At the same time, the associated
scalar curvature describes the global nature of underlying
statistical systems. Thus, it should not be difficult to establish
the behavior of state-space correlations, about an equilibrium
statistical configuration, as an ensemble of CFT microstates.
Intriguingly, we envisage that the set of parameters would form a
co-ordinate chart for the concerned intrinsic state-space
manifold.

One finds, in this spirit, that the state-space geometries
associated with the bubbling black brane solutions become rather
hard to analyze, when the number of distinct dipole charges
becomes large enough. This observation follows further from the
well-known LLM construction \cite{llm}, i.e. that in this limit
the underlying configurations have no obvious brane
interpretations, and thus there may not be an illusive set of
parameters which define their state-space configuration. More
generally, the state-space outcomes are potentially interesting,
for analyzing relevant structures of the vacuum manifold
associated with the supersymmetric vacuum states in string theory.
This seems to be based on the fact that proving or disproving the
strong form of Mathur's conjecture \cite{fuzzball}, which states
that black hole microstates are dual to smooth supergravity
solutions, finds stunning consequences and gives rise to our
state-space geometries, in the limit of Gaussian approximations.

It known that the Mathur's conjecture reduces to a definite
well-defined mathematical problem of classifying and counting
asymptotically flat four dimensional hyper K\"ahler manifolds,
which have moduli regions of signature $(+,+,+,+)$ and
$(-,-,-,-)$, see for instance \cite{Bena:2005va}. Note further,
from the existence of a weak form of the Mathur conjecture
\cite{mathurdual, fuzzball}, which states that the black hole
microstates are dual to string theory configurations with unitary
scattering, that are not necessarily smooth in the supergravity
approximation, may provide a better understanding of the
state-space geometric objects, and thus the statistical
fluctuations in them. Thus, the Mathur's conjecture, if correct,
indicates that the black holes can be contemplated as an ensemble
of hyper-K\"ahler geometries, involving foams of a very large
number of topologically non-trivial $2$-spheres, threaded by
fluxes. Hence, our state-space geometric investigation seems to
find several microscopic interpretations. Interesting state-space
inter-relations and their K\"ahler geometric or moduli space
details are nonetheless beyond the scope of the present set-up,
and thus are the matter of future explorations.
\subsection*{(iv) Generalized Hyper-K\"ahler Manifolds}
It is worth stressing that the possibility of negative
determinants of state-space geometries is not really a serious
problem, and in fact the positive definite condition is, as such,
not mandatory. This is because the signature of the base-space
manifold is allowed to change in the presence of non-zero dipole
fluxes, so that it should give rise to smooth three charge
space-time geometries. This fact may further be explained on the
basis of the classification of generalized hyper-K\"ahler
manifolds, as well. Nevertheless, we do not find it necessary to
explicate it here, for the purpose of state-space geometry, and
for developing some interesting physics of black holes/ black
strings/ black rings.

Furthermore, the black hole microstates may be described by a set
of foams of non-trivial $S^2$'s over the four-dimensional base
manifold \cite{Bena:2005va}, so one may be able to carry out some
statistical analysis, as well, for such foams, perhaps by using a
toric geometry, to see if one can describe macroscopic, bulk
state-space pair correlation functions and correlation volume of
these black brane configurations. It would also be interesting to
investigate the contributions coming from the transitions between
different base-space geometries, nucleation of GH points,
probabilities of transition, and the instanton corrections, into
our state-space manifolds. It is worth to note at this point that
the idea associated with the state-space manifolds appears to be
arising from the space-time foams, which may be important in
understanding divergence structures of the state-space/
base-space/ moduli space manifolds, as the space-time foam has
made regular appearances in the discussion of quantum gravity, see
for example \cite{HawkingZW}.

In a very general sense, what we are interrogating here is, in
somewhat similar spirit to the notions which deal with the
correlation functions in QFT or the fluctuation theory in
statistical mechanics, that they can be divulged from the fact
that the underlying brane space-times on small scales become
inevitable fuzzy topological foams \cite{Bena:2005va}. Here, we
have however managed to find state-space correlations, in the
limit of supersymmetric D-brane configurations, and with these
outcomes to a great deal, more computational control of the
problem is required, to be physically understood. Categorically,
one needs to find how the equilibrium microstates interact each
other, under the Gaussian fluctuations. These points, and their
non-equilibrium generalizations, are beyond the scope of the
present interest, and thus we leave them to open up in other
considerations.
\subsection*{(v) Physics at the Planck Scale}
It is important to note that the same physical ideas, which led to
an apprehension that the space-time becomes foamy near the Planck
scale, also come into an interplay in the state-space
considerations. This is because, at the Planck scale, there are
virtual black holes even in empty space, and we are thus analyzing
the underlying statistical correlations among the microstates,
which may be described by the aforementioned foam of two-spheres.
The concerned connotation of thermodynamic state-space geometry
may be indicated from \cite{RuppeinerPRD78,Bekenstein}.
Consistency requirements would therefore suggest that the local
and global state-space correlations of existing virtual black
holes should really be virtual fluctuations in bubbling
hyper-K\"ahler geometries. Therefore, even empty space would have
definite state-space interactions, which we have been exploring
here Thus, they should apparently have some generalization, in
terms of the parameters of foamy geometries.

Obviously, the state-space geometric description will naturally
break down at the Planck scale, but the picture is expected to
remain rather interesting, and still it is certainly supported by
the fact that large bubbles \cite{benakraus,bena,llm} are needed
to resolve singularities persisting in the macroscopic black brane
configurations. There are evidently many things to be tested and
lots of interesting things might be done, but we believe that we
have made definite important progress, by divulging the parametric
state-space manifolds and, thereby, have given an intriguing
description to local and global state-space correlations existing
among black hole equilibrium microstates. The present exploration
thus opens an avenue and may give new insight into the promising
structures of black brane space-time on very small scales.
\section*{Acknowledgement:}
\noindent The work of S. B. has been supported in part by the European Research Council grant
n. 226455, \textit{``SUPERSYMMETRY, QUANTUM GRAVITY AND GAUGE FIELDS (SUPERFIELDS)''}.
%, and
% by the European Community Human Potential Program under contract MRTN-CT-2004-005104
% \textit{``Constituents, fundamental forces and symmetries of the universe''}.
B.N.T. would like to thank Prof. R. Emparan, Prof. A. Sen and
Prof. J. de Boer for useful discussion on the stability of
rotating brane configurations and phase transitions during the
\textit{``Spring School on Superstring Theory and Related
Topics-2008, ICTP Trieste,Italy''}; Prof. J. Simon for exciting
discussions on BPS configurations and their microscopic models
during the \textit{``School on Attractor Mechanism, June-July 2009
(SAM2009), INFN, Frascati, Italy''}; Prof. V. Ravishankar for
providing viable support during the preparation of this
manuscript; Dr. B. Bhattacharjya and Dr. V. Chandra for a set of
discussions on the phases of intrinsic Riemannian and
thermodynamic geometries, while this manuscript was under
preparation. B.N.T. also acknowledges \textit{``CSIR, New Delhi,
India''}, for the financial support under the research grant
\textit{``CSIR-SRF-9/92(343)/2004-EMR-I''} and the international
travel supports of the \textit{``Indian Institute of
Technology-Kanpur, Kanpur-208016, India''} towards the
participation in the \textit{``STRINGS-2009, Roma''} and
\textit{``SAM2009 at Frascati, Italy''}, where the part of this
work was done.
%\newpage

\end{document}